\documentclass[sn-mathphys]{sn-jnlAMS}

\jyear{2021}%

\theoremstyle{thmstyleone}%
\theoremstyle{thmstyletwo}%

\theoremstyle{thmstylethree}%

\usepackage{bm,amsmath,empheq,amssymb,graphicx,stmaryrd,float,subcaption,upgreek,sidecap,dsfont,mathtools,gensymb,mathrsfs}
\usepackage[abs]{overpic}
\usepackage{hyperref,bookmark}
\usepackage{url}
\usepackage{comment,footnote}

\newcommand{\nigh}[1]{{\color{black}#1}}

\usepackage{xcolor}

\renewcommand{\br}{{\bm r}}
\newcommand{\bzero}{{\bm 0}}

\newcommand{\bx}{{\bm x}}
\newcommand{\be}{{\bm e}}
\newcommand{\by}{{\bm y}}
\newcommand{\bd}{{\bm d}}
\newcommand{\bu}{{\bm u}}

\newcommand{\bn}{{\bm n}}

\newcommand{\bI}{\mathbf{I}}

\newcommand{\bF}{{\bf F}}
\newcommand{\bL}{{\bf L}}

\newcommand{\bq}{{\bm q}}

\newcommand{\ba}{\bm{a}}
\newcommand{\bb}{{\bm b}}
\newcommand{\bnu}{{\bm \nu}}
\newcommand{\tu}{{\tilde u}}
\newcommand{\tvartheta}{{\tilde \vartheta}}
\newcommand{\tx}{{\tilde x}}
\newcommand{\tv}{{\tilde v}}
\newcommand{\tLambda}{{\tilde \Lambda}}

\newcommand{\ave}[1]{{\left\langle #1 \right\rangle}}
\newcommand{\trans}{^\mathsf{T}}

\newcommand{\n}{\bm{n}}
\newcommand{\m}{\bm{n}_0}
\renewcommand{\e}{\bm{e}}

\renewcommand{\x}{\bm{x}}

\renewcommand{\y}{\bm{y}}
\newcommand{\normal}{\bm{\nu}}

\newcommand{\surface}{\mathscr{S}}
\newcommand{\free}{\mathscr{F}}
\newcommand{\nablas}{\nabla\!_{\mathrm{s}}}
\renewcommand{\f}{\bm{f}}

\newcommand{\F}{\mathbf{F}}
\renewcommand{\C}{\mathbf{C}}
\newcommand{\Cf}{\C_{\f}}
\newcommand{\nay}{(\nabla\y)}
\newcommand{\curvature}{(\nablas\normal)}

\newcommand{\ribbon}{\mathscr{S}_0}
\newcommand{\slab}{\mathsf{S}}

\newcommand{\Lc}{\mathbf{L}}
\newcommand{\Lr}{\mathbf{L}_0}
\newcommand{\sphere}{\mathbb{S}^2}

\DeclareMathOperator{\tr}{tr}
\usepackage[utf8]{inputenc}
\usepackage[T1]{fontenc}
\usepackage{nomencl}

\begin{document}

\title[Bending and Stretching of a NPN Ribbon]{Bending and Stretching in a Narrow Ribbon of Nematic Polymer Networks}


\author[1]{\fnm{Harmeet} \sur{Singh}}\email{harmeet.singh@epfl.ch}

\author*[2]{\fnm{Epifanio G.} \sur{Virga}}\email{eg.virga@unipv.it}
\equalcont{These authors contributed equally to this work.}

\affil[1]{\orgdiv{Laboratory for Computation and Visualization in Mathematics and Mechanics, Institute of Mathematics}, \orgname{\'Ecole Polytechnique F\'ed\'erale de Lausanne}, \orgaddress{\street{} \city{Lausanne}, \postcode{CH-1015}, \country{Switzerland}}}

\affil*[2]{\orgdiv{Dipartimento di Matematica}, \orgname{Universit\`a di Pavia}, \orgaddress{\street{via Ferrata 5}, \city{Pavia}, \postcode{I-27100}, \country{Italy}}}


\abstract{	We study the spontaneous out-of-plane bending of a planar untwisted ribbon composed of nematic polymer networks activated by a change in temperature. Our theory accounts for both stretching and bending energies, which compete to establish equilibrium. We show that when equilibrium is attained  these energy components obey a \emph{complementarity} relation: one is maximum where the other is minimum. Moreover, we identify a \emph{bleaching} regime: for sufficiently large values of an activation parameter (which measures the mismatch between the degrees of order in polymer organization in the reference and current configurations), the ribbon's deformation is essentially independent of its thickness.}

\keywords{Nematic polymer networks, Nematic glasses, Soft matter elasticity, Nematic elastomers, Ribbon theory, Thermally activated elastic materials}


\pacs[MSC Classification]{Primary 74, Secondary 74B20, 74K10, 74K35, 76A15}

\maketitle

\section{Introduction}\label{sec:intro}
A nematic elastomer differs from a standard elastomer in that it hosts \emph{nematogenic} rod-like molecules within its polymer strands. If the temperature is sufficiently low, the nematogenic components develop a tendency to be aligned alike (the very signature of a nematic liquid crystal), thus introducing a degree of anisotropy in the spatial organization of polymer strands. This suffices to alter the elastic response of these materials. Since the degree of nematic order depends on temperature, a change in the latter can induce a spontaneous deformation of the body.


The nematic order is described by a scalar, here denoted $Q$,  and a unit vector $\n$ (or \emph{director}), where the former has a bearing on the spatial organization of polymer strands and the latter represents the average orientation of the nematogenic constituents.

Nematic elastomers can be envisioned as a fluid-solid mixture, where the nematic component is fluid and the cross-linked polymer matrix is solid \cite{corbett:photomechanics}. Generally, fluid and solid components enjoy a certain degree of mutual independence. A full spectrum of materials exist, whose mechanical behavior can be classified in accordance with the degree of freedom allowed to $\n$ within the polymer matrix \cite{white:photomechanical_collection,ware:programmed}.

\emph{Nematic polymer networks} (NPNs) are nematic elastomers of a special kind: their cross-linking is so tight that the nematic director is completely \emph{enslaved} to the matrix deformation. Only these materials will form the object of our study.

We shall assume that a NPN is cross-linked in the nematic phase, which will constitute the stress-free, reference configuration of the bodies considered here. Deformations will be induced by a change in temperature, which drives the reference configuration out of equilibrium.\footnote{A change in temperature affects the degree of orientational order among nematogenic moulecules; its effect is similar to that driven at constant temperature by the shape change induced by light on photoresponsive nematic molecules (not involved here). See, for example, \cite{sonnet:model} for a discussion on photoactivable NPNs.} A scalar activation parameter will describe the mismatch between nematic order in the reference and current configurations; it will act as a control parameter.

We are interested in the slimiest of bodies, a \emph{ribbon}. Our continuum theory is based on the three-dimensional theory of nematic elastomers revolving around the celebrated \emph{trace formula} for the volume elastic energy density, which has long been studied  \cite{blandon:deformation,warner:theory,warner:elasticity,warner:nematic} (as also illustrated in Sect.~6.1 of \cite{warner:liquid} under the suggestive heading of \emph{neo-classical} theory).

For the readers interested in broadening their background on nematic elastomers who are intimidated by the vast available literature, a recommended reference is  the influential book by Warner and Terentjev~\cite{warner:liquid}; the theoretical literature that preceded and prepared for it \cite{blandon:deformation,warner:soft,terentjev:orientation,verwey:soft,verwey:multistage,verwey:elastic} is also of interest. General continuum theories have also been proposed \cite{anderson:continuum,zhang:continuum,mihai:nematic}, some also very recently. Applications are plenty; a collection can be found in a book \cite{white:photomechanical_book} and a recent special issue \cite{korley:introduction}. Finally, some valuable guidance can be gained from the reviews \cite{mahimwalla:azobenzene,ube:photomobile,white:photomechanical,ula:liquid,pang:photodeformable,kuenstler:light,warner:topographic}. 

This paper is an outgrowth of a previous study \cite{singh:model}, where we derived our theory for NPN ribbons from a reduction procedure from three to two space dimensions proposed in \cite{ozenda:blend} with the intent of enclosing both stretching and bending contents in the surface elastic energy of thin NPN sheets. To make our development self-contained, we recall in Sect.~\ref{sec:preliminaries} the fundamentals of our theory; the articulation in subsections will aid the experienced reader to gather the essential information. In Sect.~\ref{sec:narrow_rectangular_ribbon}, we specify the kinematics of  the simple problem that we consider here. Section~\ref{sec:variational_problem}, which is the heart of the paper, is where we state and solve the variational problem for the optimal shape of a thermally activated ribbon. A plethora of activated equilibria are described in Sect.~\ref{sec:results} for a whole range of activation and geometric parameters. Finally, in Sect.~\ref{sec:conclusions}, we collect two qualitative conclusions of our study: one identifies a regular pattern in the relative distribution of stretching and bending energies at equilibrium, the other suggests the existence of an activation regime where the ribbon's deformation is insensitive to its thickness. 

\section{Preliminaries}\label{sec:preliminaries}
To make our paper self-contained, we summarize in this section some results already obtained in \cite{ozenda:blend} and \cite{singh:model}. The reader familiar with this material may jump ahead to the following sections.

Our theory is based on the \emph{trace formula} for the elastic volume energy density (in the reference configuration) of nematic elastomers in the form put forward by \cite{finkelmann:elastic},
\begin{equation}
	\label{eq:free_energy_density}
	f_\mathrm{e}=\frac12k\bigg\{\tr(\F\trans\Lc^{-1}\F\Lr)+\ln\left(\frac{\det\Lc}{\det\Lr}\right)\bigg\},
\end{equation}
where $\F$ is the deformation gradient and the \emph{shear} elastic modulus $k$ is given by
\begin{equation}
	\label{eq:mu_0}
	k=n_\mathrm{s}k_BT,
\end{equation}
in terms of the number density of polymer strands $n_\mathrm{s}$, the Boltzmann constant $k_B$, and the absolute current temperature $T$.
In \eqref{eq:free_energy_density}, $\Lr$ and $\Lc$ are the \emph{step-length} tensors of the material in the reference and current configurations, respectively. They are generally different because different are the arrangements of polymer chains in the two configurations, if a change in temperature has occurred, altering the orientational order of the nematogenic molecules that comprise the polymer strands. 
Following \cite{verwey:elastic,nguyen:theory}, we represent them as 
\begin{equation}
	\label{eq:step_tensor}
	\bL_0=A_0(\bI+S_0\bn_0\otimes\bn_0)\quad\text{and}\quad\bL=A(\bI+S\bn\otimes\bn),
\end{equation}
where $\bI$ is the identity (in three-dimensional space) and $\bn_0$ and $\n$ are the nematic directors in reference and current configurations. Moreover,
\begin{subequations}
	\label{eq:As_and_Ss}
	\begin{equation}
		\label{eq:A_0_S_0}
		A_0=\ell_\mathrm{m}(1-Q_0),\quad S_0=\frac{3Q_0}{1-Q_0}
	\end{equation}
	and their twins,
	\begin{equation}
		\label{eq:A_S}
		A=\ell_\mathrm{m}(1-Q),\quad S=\frac{3Q}{1-Q},
	\end{equation}
\end{subequations}
are  expressions derived within the classical statistical mechanics model for a polymer chain thought of  as consisting in freely jointed nematogenic monomers, each of length $\ell_\mathrm{m}$ (see Chapt.~6 of \cite{warner:liquid} and also \cite{corbett:nonlinear,corbett:polarization}). In \eqref{eq:As_and_Ss}, $Q_0$ (and $Q$) represents the Maier-Saupe nematic scalar order parameter describing the degree of orientation of monomers in the reference (and current) configuration of the material.\footnote{This theory is an extension of the classical Gaussian rubber elasticity. An exposition of statistical theories for rubber can be found in the reference book \cite{treloar:non-linear_third}. Adaptation of the simplest realization of these theories \cite{deam:theory} to the case where the distribution of monomers in a polymer chain is anisotropic delivers \eqref{eq:free_energy_density}.}

$Q$ (and, correspondingly, $Q_0$) is defined as 
\begin{equation}
	\label{eq:Q_definition}
	Q:=\ave{P_2(\n\cdot\bu)},
\end{equation}
where $P_2$ is the second Legendre polynomial, $\bu\in\sphere$ is the unit vector along a single monomer, and the brackets $\ave{\cdots}$ denote ensemble average. $Q$ ranges in the interval $[-\frac12,1]$, whose end-values represent the limiting cases of $\bu$'s distributed isotropically in the plane orthogonal to $\n$ and $\bu$'s aligned with $\n$, respectively. 

Here, we shall assume that $Q$ (and $Q_0$) ranges in $(0,1)$. In this interval, by \eqref{eq:A_S}, $S$ is a monotonically increasing function of $Q$, which also increases when the temperature $T$ is decreased below the isotropic-to-nematic transition. Thus, a decrease in temperature results in an increase in $S$. This explains why the mechanism of thermal activation can be effectively described as having $S\neq S_0$. A temperature increase makes $S<S_0$, whereas a temperature decrease makes $S>S_0$.

When, as in our case,  the scalar order parameters $S_0$ and $S$ are prescribed, the second term in \eqref{eq:free_energy_density} is not affected by the deformation and can be omitted, thus reducing \eqref{eq:free_energy_density} to the following \emph{bare} trace formula of \cite{blandon:deformation} (also discussed in \cite{warner:new}), which depends only on $\F$ and $\n$ and will be adopted from now on,
\begin{equation}
	\label{eq:f_e}
	f_\mathrm{e}=\frac12k\tr(\bF\trans\bL^{-1}\bF\bL_0).
\end{equation}     

The degree of cross-linking in the material is responsible for the mobility of the nematic director relative to the network matrix \cite{ware:programmed}. A \emph{nematic polymer network} (NPN) represents the tightest end of the cross-linking spectrum, where $\n$ is \emph{enslaved} to the deformation.\footnote{The name \emph{nematic polymer network} was proposed in \cite{white:programmable}. Other authors prefer to call these materials \emph{nematic glasses} (see, for example, \cite{modes:disclination}).} For these materials, which are the only ones considered here, we assume that 
\begin{equation}\label{eq:n_conveyed}
	\n=\frac{\F\m}{\vert\F\m\vert},
\end{equation}
which says that the nematic director $\m$ imprinted in the polymer network at the cross-linking time is \emph{conveyed} by the solid matrix of the body. We further assume that the material is \emph{incompressible},\footnote{Even if, as discussed in \cite{white:photomechanical_collection}, not all NPNs are strictly so.} so that $\F$ is subject to the constraint
\begin{equation}
	\label{eq:incompressibility}
	\det\F=1.
\end{equation}

By use of \eqref{eq:step_tensor}, \eqref{eq:mu_0}, \eqref{eq:A_S}, and \eqref{eq:n_conveyed},  \eqref{eq:f_e} can be given the following form \cite{ozenda:blend},
\begin{equation}\label{eq:F_definition}
	f_\mathrm{e}=\frac12e_0E(\Cf),
\end{equation} 
where
\begin{equation}
	\label{eq:e_0_definition}
	e_0:=k\frac{A_0}{A}=n_\mathrm{s}k_BT\frac{3+S}{3+S_0},
\end{equation}
$\Cf:=\bF\trans\bF$ is the right Cauchy-Green tensor associated with the deformation $\f$ and 
\begin{equation}\label{eq:bulk_energy_density}
	E(\Cf):=\tr\Cf+\frac{S_0}{S+1}\n_0\cdot\Cf\n_0-\frac{S}{S+1}\frac{\n_0\cdot\Cf^2\n_0}{\n_0\cdot\Cf\n_0}.
\end{equation}

To apply  our theory to the spontaneous  deformation of a ribbon out of its plane (which is the theme of the following sections), we must first learn how to reduce the volume free-energy density in \eqref{eq:F_definition} to a surface free-energy density to be attributed to a thin sheet, thought of as a two-dimensional body. This task was accomplished in \cite{ozenda:blend}; in the following subsection, we recall the main results of this study.

\subsection{Dimension reduction}\label{sec:dimension_reduction}
The method adopted in \cite{ozenda:blend} is classical and consists in performing an expansion of $f_\mathrm{e}$ retaining up to the cubic terms in the sheet's thickness, thus identifying \emph{stretching} and \emph{bending} contents in the surface energy by the power in the thickness they scale with.

\begin{figure}[h]
	\centering
	\includegraphics[width=.5\linewidth]{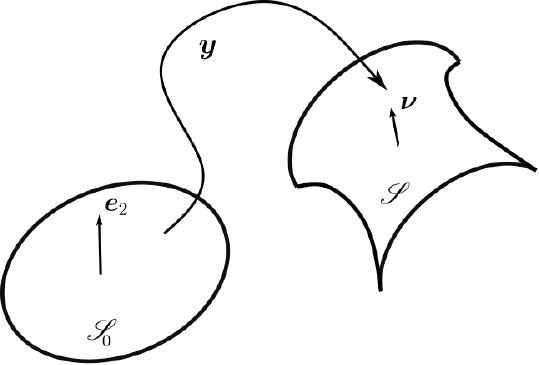}
	\caption{Schematic representation of the deformation of a thin sheet. $\surface_0$ is the planar midsurface of the slab $\slab$ of thickness $2h$ representing the reference configuration of the body; it lies in the plane $\{\e_3,\e_1\}$ of Cartesian frame $\{\e_1,\e_2,\e_3\}$. $\surface$ is the image under the mapping $\y$ of $\surface_0$; it is an oriented surface with unit normal $\normal$.}
	\label{fig:dimension_reduction}
\end{figure}

We identify the undeformed body with a slab $\slab$ of thickness $2h$ and midsurface $\surface_0$ in the $\{\e_3,\e_1\}$ plane of a Cartesian frame. We further assume that $\m$ is imprinted in $\slab$ so that it does not depend on the $x_2$ (out-of-plane) coordinate and  $\m\cdot\e_2\equiv0$. Moreover, we represent the three-dimensional deformation $\f$ as
\begin{equation}
	\label{eq:deformation_representation}
	\f(\x,x_2)=\y(\x)+\phi(\x,x_2)\normal,
\end{equation} 
where $\x$ varies in $\surface_0$, $x_2$ ranges in the interval $[-h,h]$, $\normal$ is the normal to the midsurface $\surface=\y(\surface_0)$ of the deformed slab $\f(\slab)$ (see Fig.~\ref{fig:dimension_reduction}),
and $\phi(\x,x_2)$ is a function to be determined,\footnote{In the classical theory of plates, the \emph{Kirchhoff-Love} hypothesis stipulates that $\phi\equiv x_2$ (see, for example, \cite{podio:exact}, for a modern treatment). In \cite{ozenda:kirchhoff}, the Kirchhoff-Love hypothesis was reformulated in the more general form adopted here and criteria were suggested to identify the function  $\phi$, none of which delivered exactly the original Kirchhoff-Love form.} enjoying the property
\begin{equation}\label{eq:Phi_property}
	\phi(\x,0)=0.	
\end{equation}	

As shown in \cite{ozenda:blend}, the constraint of incompressibility \eqref{eq:incompressibility} determines $\phi$ in the form
\begin{equation}\label{eq:Phi_determination}
	\phi(\x,x_2)=x_2-Hx_2^2+\frac13(6H^2-K)x_2^3 +O(x_2^4),
\end{equation}
where $H$ and $K$ are the \emph{mean} and \emph{Gaussian} curvature of $\surface$, respectively, defined as
\begin{equation}
	\label{eq:H_and_K}
	H(\y(\x)):=\frac12\tr\curvature\quad\text{and}\quad K(\y(\x)):=\det\curvature,
\end{equation}
in terms of the two-dimensional curvature tensor $\nablas\normal$ at the point $\y(\x)$ on $\surface$.

Moreover, under assumption \eqref{eq:n_conveyed} and for $h$ sufficiently small, the nematic director $\n$ in the present configuration $\f(\slab)$ is delivered by (see \cite{sonnet:model})
\begin{equation}
	\label{eq:n_present_configuration}
	\n(\f(\x,x_2))=\frac{(\bI+\phi\nablas\normal)\nay\m}{\lvert(\bI+\phi\nablas\normal)\nay\m\rvert}.
\end{equation}
Since $\m\cdot\e_2\equiv0$, $\nay\m\cdot\normal\equiv0$, and $\nablas\normal$ is a symmetric tensor mapping the local tangent plane to $\surface$ into itself, \eqref{eq:n_present_configuration} shows that $\n\cdot\normal=0$ everywhere within $\f(\slab)$, but while $\m$ is \emph{uniform} across the thickness of $\slab$, $\n$ is \emph{not} so across the thickness of $\f(\slab)$. However, by \eqref{eq:n_present_configuration} and \eqref{eq:Phi_property}, on $\surface$ $\n$ is given by
\begin{equation}
	\label{eq:y_inextensibility}
	\n(\f(\x,0))=\frac{\nay\m}{\lvert\nay\m\rvert},
\end{equation} 
and so it appears to be conveyed by the deformation of $\surface_0$, in accordance with the three-dimensional constraint \eqref{eq:n_conveyed}. Accordingly, to mimic \eqref{eq:incompressibility}, we shall assume that 
\begin{equation}
	\label{eq:det_nabla_y}
	\det\nay=1,
\end{equation}
which makes $\surface_0$ an \emph{inextensible} surface.

Laborious calculations, building upon  \eqref{eq:Phi_determination}, \eqref{eq:y_inextensibility}, and \eqref{eq:det_nabla_y}, established in \cite{ozenda:blend} that the surface elastic energy density for the deformation of $\surface_0$ into $\surface$ is given by
\begin{equation}
	\label{eq:f_e_tilde}
	\widetilde{f}_\mathrm{e}=\frac12e_0h(f_\mathrm{s}+f_\mathrm{b}) +O(h^5),
\end{equation}
where 
\begin{subequations}\label{eq:f_s_and_f_b}
	\begin{eqnarray}
		f_\mathrm{s}&=&\frac{2}{\nigh{S}+1}\left(\tr\C+S_0\n_0\cdot\C\n_0+\frac{S}{\n_0\cdot\C\n_0}\right),\label{eq:f_s}\\
		f_\mathrm{b}&=&\frac{2h^2}{3}\bigg\{2(8H^2-K)\nonumber\\&+&\frac{1}{S+1}\left[\left(\frac{3S}{a_0^2}-a_0^2S_0-\tr\C\right)K-\frac{4S}{a_0^2}(2H-\kappa_n)\kappa_n\right] \bigg\}\label{eq:f_b}
	\end{eqnarray}
\end{subequations}
are the (dimensionless) stretching and bending free-energy \emph{contents}, respectively. In \eqref{eq:f_s_and_f_b}, $\C:=\nay\trans\nay$ is the right Cauchy-Green tensor associated with the deformation $\y$, $a_0^2 := \n_0\cdot\C\n_0$, and
\begin{equation}
	\label{eq:11}
	\kappa_n:=\n\cdot\curvature\n,
\end{equation}
which couples $\n$ with the geometry of $\surface$.

As shown in detail in Appendix~\ref{sec:stretching}, in the absence of activation, that is, for $S=S_0$, $f_\mathrm{s}$ attains its minimum subject to \eqref{eq:det_nabla_y} for $\C=\mathbf{I}_2$, where $\mathbf{I}_2$ is the identity in two-dimensional space. Thus, in the membrane approximation, the surface elastic energy of an \emph{inactive} sheet is minimized by isometries.

Finally, dropping in \eqref{eq:f_e_tilde} the higher order terms in $h$, we write the total elastic free-energy functional in the form
\begin{equation}
	\label{eq:free_energy_functional}
	\free[\y]:=\frac12e_0h\int_{\surface_0}(f_\mathrm{s}+f_\mathrm{b})dA,
\end{equation}
where now $A$ denotes the area measure and $e_0h$ is a characteristic surface energy density precisely as $e_0$ in \eqref{eq:e_0_definition} is a characteristic volume energy density.

\subsection{General theory of NPN ribbons}\label{sec:general}
We reproduce here only the essential aspects of the theory of NPN ribbons, obtained through a dimension reduction procedure developed in \cite{singh:model}.
The procedure reduces the surface energy density of a sheet of NPNs, derived in \cite{ozenda:blend}  and summarized in Sect.~\ref{sec:preliminaries}, to a line energy density valid for ribbons, whose reference and deformed configurations are assumed to be ruled surfaces.

\begin{figure}[h]
	\centering
	\includegraphics[width=\textwidth]{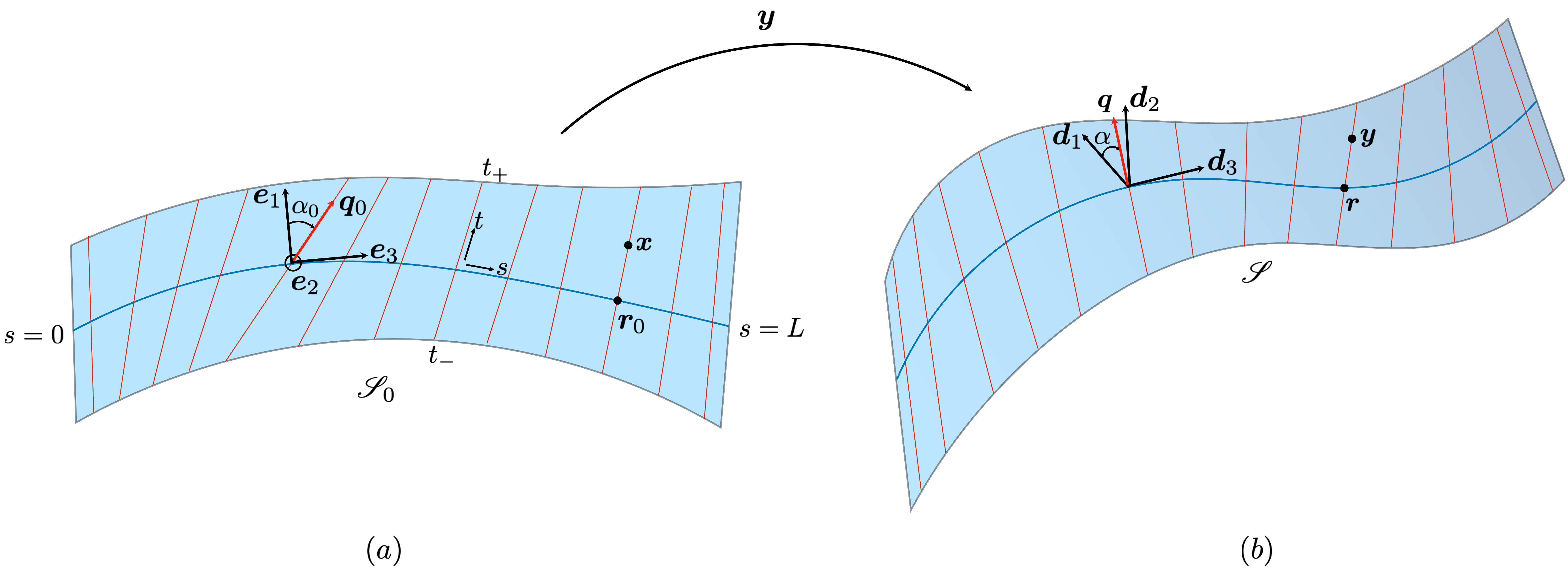}
	\caption{(a) The reference configuration $\ribbon$ of a planar ribbon, ruled by the directrix $\br_0(s)$ and the generators in the direction of the unit vector $\bq_0(s)$. An orthonormal set of directors $\{\be_1(s),\be_2,\be_3(s)\}$ are adapted to the centreline such that $\be_3(s)$ is tangential to $\br_0(s)$, and $\be_2$ is pointing out of the plane of the paper. The angle between $\bq_0$ and $\be_1$ is denoted by $\alpha_0$. (b) A current, non-planar, configuration $\mathscr{S}$ of the ribbon with the directrix denoted by $\br(s)$, and the direction of the generators given by $\bq(s)$. An orthonormal triad $\{\bd_1(s),\bd_2(s),\bd_3(s)\}$ is attached to the centreline with $\bd_3(s)$ oriented along the tangent vector, and $\bd_1(s)$  in the tangent plane at $\surface$. The angle between $\bd_1(s)$ and $\bq_0(s)$ is shown by $\alpha$.}
	\label{fig:general_schematics}
\end{figure}

Consider a ribbon made of NPNs, whose stress-free natural configuration $\mathscr{S}_0$, and deformed configuration $\mathscr{S}$ are depicted in Fig.~\ref{fig:general_schematics}.
We should think of them as follows. $\surface_0$ is a stress-free configuration of the inactive ribbon (when $S=S_0$). Once activated, so that $S\neq S_0$, $\surface_0$ is no longer stress-free and the ribbon falls out of equilibrium; a deformation then restores equilibrium by morphing $\surface_0$ into $\surface$.

We choose a material line $\br_0\in\mathscr{S}_0$ in the natural configuration, endowed with a triad of orthonormal vectors $\{\be_1,\be_2,\be_3\}$, oriented as shown in Fig.~\ref{fig:general_schematics}. Here $\e_2$ is independent of $s$, while $\{\e_3,\e_1\}$ is a movable frame in the plane orthogonal to $\e_2$. Letting $\bm{\omega}$ denote the Darboux vector associated with the frame $\{\be_1,\be_2,\be_3\}$, so that $\partial_s\e_i=\bm{\omega}\times\e_i$, we easily see that $\bm{\omega}=\omega_2\e_2$.
We refer to $\br_0$ as the directrix, or the centerline, of the ribbon.
We parameterize the natural configuration as a ruled surface with the following representation,
\begin{equation}
	\bx(s,t) = \br_0(s) + t\bq_0(s)\, ,\label{eq:x}
\end{equation}
where $\bq_0(s)$ is a unit vector given by $\bq_0(s) = \cos\alpha_0(s)\be_1 + \sin\alpha_0(s)\be_3$ with $-\frac{\pi}{2}\le\alpha_0\le\frac{\pi}{2}$.
Here $s$ and $t$ are material coordinates along the centerline $\br_0$ and $\bq_0$, respectively, and $\alpha_0$ is a smooth function of $s$ such that $\bq_0(0)$ and $\bq_0(L)$ coincide, respectively, with the short edges of the ribbon at $s=0$ and $s=L$. Formally, in the $(s,t)$ coordinates, the ribbon is represented by the set $\{(s,t):0\leqq s\leqq L, t_-(s)\leqq t\leqq t_+(s)\}$, where $t_\pm(s)$ are functions describing the two long sides of the ribbon.
We will identify the unit vector $\bq_0$ with the imprinted nematic director $\bn_0$, thereby imparting material character to the former,
\begin{equation}\label{eq:n_0_identification}
	\bq_0 = \bn_0\, .
\end{equation}

When the ribbon deforms, the centerline $\br_0(s)$ and the unit vectors $\bq_0(s)$ are convected to $\br(s)\in\mathscr{S}$ and $\bq(s)\in\mathscr{S}$, where $\mathscr{S}$ denotes a deformed configuration of the ribbon.
Similar to $\br_0(s)$, the deformed centerline $\br(s)$ is endowed with an ordered orthonormal director frame $\{\bd_1(s),\bd_2(s),\bd_3(s)\}$, where $\bd_3(s)$ is constrained to lie along the tangent (see Fig.~\ref{fig:general_schematics}).
The kinematics of the centerline are captured by the following two relations,
\begin{equation}
	\partial_s\br = v_3\bd_3\quad\text{and}\quad\partial_s\bd_i = \bu\times\bd_i\, ,\label{eq:centerline_kinematics}
\end{equation}
where $v_3(s)$ is identified as the stretch of the centerline, and $\bu(s)$ is the Darboux vector associated with the frame $\{\bd_1,\bd_2,\bd_3\}$. The director components in the expansion $\bu=u_1\bd_1+u_2\bd_2+u_3\bd_3$ represent the bending strains about the corresponding directors.

The deformed configuration $\mathscr{S}$ of the ribbon is parameterized as
\begin{equation}
	\by(s,t) = \br(s) + \psi(s,t)\bq(s)\, ,\label{eq:y}
\end{equation}
where $\bq(s)$ is a unit vector given by
\begin{equation}
	\label{eq:q_representation}
	\bq(s) = \cos\alpha(s)\bd_1(s) + \sin\alpha(s)\bd_3(s)\quad\text{with}-\frac{\pi}{2}\le\alpha\le\frac{\pi}{2}
\end{equation}
and $\psi(s,t)$ is a smooth function. Letting $\bq_0^\perp$ be the unit vector in the $\{\e_3,\e_1\}$ plane defined by
\begin{equation}
	\label{eq:q_0_perp_definition}
	\bq_0^\perp:=\e_2\times\bq_0,
\end{equation}
we represent the deformation gradient $\nabla\y$ as
\begin{equation}
	\nabla\by = \ba\otimes\bq_0 + \bb \otimes\bq_0^\perp\, ,\label{eq:deformation_gradient}
\end{equation}
where $\ba$ and $\bb$ are the images of $\bq_0$ and $\bq_0^\perp$ in the current configuration. In the present setting, \eqref{eq:det_nabla_y} amounts to the condition
\begin{equation}
	\label{eq:a_times_b}
	\vert\ba\times\bb\vert=1.
\end{equation}
It was proved in \cite{singh:model} that \eqref{eq:a_times_b} is equivalently represented by the following relations,
\begin{subequations}\label{eq:inextensibility_solutions}
	\begin{align}
		\psi &= a(s)t\quad\text{with}\quad a(s)\ge\frac{\cos\alpha_0}{v_3}>0\, ,\label{eq:psi}\\
		\alpha &=\arccos\left(\frac{\cos\alpha_0}{v_3 a}\right)\, ,\label{eq:alpha}\\
		u_2 &=\partial_s\alpha - \frac{\partial_s\alpha_0}{a^2}\, ,\label{eq:u2}\\
		u_3&=u_1\tan\alpha\, ,\label{eq:u_3}
	\end{align}
\end{subequations}
where $a$ and $v_3$ are independent deformation measures.

Equations \eqref{eq:inextensibility_solutions} have important consequences (see again \cite{singh:model}), namely,
\begin{equation}\label{eq:consequence_1}
	\partial_s\bq = -\frac{\partial_s\alpha_0-\omega_2 }{a^2} \bq^\perp\, ,
\end{equation}
where $\bq^\perp:=\bd_2\times\bq$, and
\begin{subequations}
	\label{eq:consequence_2}
	\begin{align}
		\ba&=a\bq,\label{eq:a_b_solutions_a}\\
		\bb&=-\left[\frac{v_3\sin\alpha-a\sin\alpha_0+t \partial_s a }{\cos\alpha_0+t(\partial_s\alpha_0-\omega_2)}\right]\bq + \frac{1}{a}\bq^\perp\, .\label{eq:a_b_solutions_b}
	\end{align}
\end{subequations}
The latter two equations express $\ba$ and $\bb$ in the orthonormal frame $\{\bq,\bq^\perp\}$ that spans the local tangent plane at $\y(\x)$ to the current configuration of the ribbon.  Moreover, in light of  \eqref{eq:a_b_solutions_a}, \eqref{eq:deformation_gradient}, and \eqref{eq:n_0_identification}, \eqref{eq:y_inextensibility} implies that 
\begin{equation}
	\label{eq:n_identification}
	\bq=\n,
\end{equation}
which confers a material character to the rulings of the surface representing via 
\eqref{eq:y} the current configuration of the ribbon.

It also follows from the inextensibility constraint \eqref{eq:a_times_b} that the unit normal $\normal$ to $\surface$ can be written as 
\begin{equation}
	\bnu(s,t)=\bq(s)\times\bq^\perp(s) = \bd_2(s)\, ,\label{eq:normal}
\end{equation}
from which we derived in \cite{singh:model} the following formula for the curvature tensor,
\begin{equation}
	\nablas\bnu=\sigma\bq^\perp\otimes\bq^\perp\, ,\label{eq:curvature_tensor}
\end{equation}
where
\begin{equation}\label{eq:curvature}
	\sigma =\frac{a^2u_1v_3}{\cos\alpha_0\left[\cos\alpha_0+t(\partial_s\alpha_0-\omega_2)\right]},
\end{equation} 
so that, in particular, the Gaussian curvature $K$ of $\surface$ vanishes identically, whereas the mean curvature $H=\sigma/2$.

In \cite{singh:model}, using the kinematics described above along with relations \eqref{eq:inextensibility_solutions}, we obtained a dimension reduction of the sheet elastic  energy in \eqref{eq:free_energy_functional} to an energy defined over the centerline of a generic ribbon with a planar natural configuration.
The resulting energy functional is given by
\begin{equation}\label{eq:elastic_free_energy}
	\free[\y] =\frac12e_0h\int_0^L\!  f\, ds,
\end{equation}
where $e_0$ is the characteristic volume energy density introduced in \eqref{eq:e_0_definition},
\begin{equation}
	f= F_0\ln\left(\frac{\cos\alpha_0 + t_+ M}{\cos\alpha_0 + t_- M}\right) +F_1 (t_+ - t_-)+F_2 (t_+^2 - t_-^2)\,,\label{eq:F}
\end{equation} 
and
\begin{subequations}
	\begin{align}
		F_0 & = \left[\frac{8 a^4 h^2 u_1^2 v_3^2}{3 M \cos\alpha_0^2}+\frac{2\left(VM - \partial_s a\cos\alpha_0\right)^2}{(1+S)M^3}\right]\, ,\label{eq:F_0}\\
		F_1 & =2 \left[\frac{\cos\alpha_0}{a^2}+\frac{1+S_0}{1+S}a^2\cos\alpha_0+\frac{\partial_s a (2VM - \partial_sa\cos\alpha_0)}{(1+S) M^2}\right]  \, ,\label{eq:F_1}\\
		F_2 &= \left[\frac{M}{a^2} + \frac{1+S_0}{1+S}a^2 M+\frac{(\partial_s a)^2}{(1+S)M}\right]\, ,\label{eq:F_2}\\
		V & = v_3\sin\alpha - a \sin\alpha_0 \, ,\label{eq:A}\\
		M &= \partial_s\alpha_0 - \omega_2\,.\label{eq:M}
	\end{align}
\end{subequations}

\section{A Narrow Rectangular Ribbon} \label{sec:narrow_rectangular_ribbon}
\begin{figure}[h]
	\centering
	\includegraphics[width=.9\linewidth]{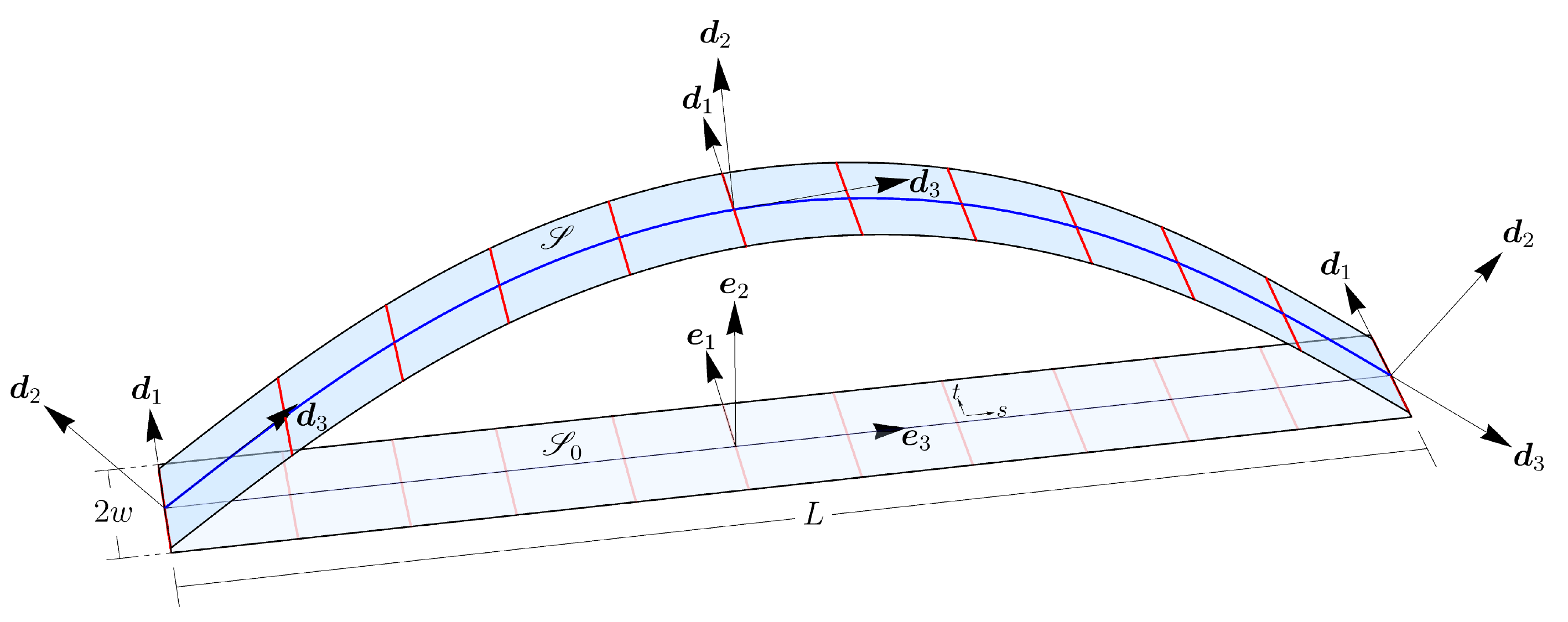}
	\caption{Schematic of  natural and deformed configurations of a narrow NPN ribbon of length $L$ and width $2w$. The width of the ribbon has been exaggerated for presentation purposes. }
	\label{fig:rectangular_schematics}
\end{figure}

Our focus in this article is on a simple ribbon with a rectangular geometry, as shown in Fig.~\ref{fig:rectangular_schematics}.
We assume the centerline $\br_0$ in the reference configuration to be a straight line such that it lies along $\be_3$, so that $\omega_2\equiv0$.
We further impose a condition on $\alpha_0$ such that, 
\begin{equation}
	\alpha_0(0) = \alpha_0(L) = 0 \, .\label{eq:alpha0}
\end{equation}
which ensures that the shorter edge of the ribbon is along $\be_1$, and that the rulings \eqref{eq:x} cover the entire material surface of the natural configuration.
As a consequence of this choice, we have,
\begin{equation}
	t_\pm =\pm\frac{w}{\cos\alpha_0}\, .\label{eq:width_t}
\end{equation}

To obtain an energy density for a \emph{narrow} ribbon, we consider the leading order term in $w$ in the expansion of \eqref{eq:F}, under which the energy functional takes the following form,
\begin{equation}
	F[\alpha,v_3] := \int_0^1\left(f_1 h + f_3 h^3\right)ds\, .\label{eq:energy}
\end{equation}
Here $F$ is renormalized by the scaling energy $2wL^2e_0$. The total length $L$ of the centerline in the reference configuration is taken to be the length scale, and consequently, all lengths from this point onward will be stated in units of $L$.
Furthermore, $f_1$ and $f_3$ are given by,
\begin{subequations}\label{eq:energy_densities}
	\begin{align}
		f_1 &=\frac{v_3^2\cos^2\alpha}{\cos^2\alpha_0}+\frac{1+S_0}{1+S}\frac{\cos^2\alpha_0}{v_3^2\cos^2\alpha}+\frac{(v_3^2\cos\alpha\sin\alpha-\cos\alpha_0\sin\alpha_0)^2}{(1+S)v_3^2\cos^2\alpha_0\cos^2\alpha}\, ,\label{eq:f1}\\
		\quad f_3 &= \frac{4}{3}\frac{u_1^2}{v_3^2\cos^4\alpha}\, .\label{eq:f3}
	\end{align}
\end{subequations}
Although $a$ has a transparent geometric meaning, defining $F$ in \eqref{eq:energy} we preferred to express it in terms of two other (independent) measures of deformation, namely, $v_3$ and the angle $\alpha$ that $\n$ makes with $\bd_1$. They are related to $a$ through the equation
\begin{equation}
	\label{eq:a_alpha}
	a=\frac{\cos\alpha_0}{v_3\cos\alpha},
\end{equation}
which makes \eqref{eq:alpha} identically satisfied.
Similarly \eqref{eq:u2}  can be reduced to the following 
\begin{equation}
	u_2=\partial_s\alpha-\partial_s\alpha_0\left(\frac{v_3\cos\alpha}{\cos\alpha_0}\right)^2.\label{eq:u2_constraint}
\end{equation}

Our objective here is to minimize $F$ in \eqref{eq:energy} under constraints \eqref{eq:u2_constraint} and \eqref{eq:u_3} with an appropriate set of boundary conditions leading to out-of-plane bending of the ribbon. $S_0$ is the scalar order parameter frozen in the reference configuration at the time of cross-linking, while $S$ is the current scalar order parameter induced by a change in temperature.\footnote{It should be recalled that they are related through \eqref{eq:As_and_Ss} to the Maier-Saupe nematic scalar order parameters $Q_0$ and $Q$ in the corresponding configurations of the material.} Since both $S$ and  temperature $T$ are held fixed as the spontaneous deformation  unfolds, the minimizers of $F$ are not affected by its being scaled to a quantity, $e_0$, depending on both $S$ and $T$.

\section{The Variational Problem}\label{sec:variational_problem}
We seek solutions with out-of-plane deformation of a narrow rectangular NPN ribbon on which the following boundary conditions are prescribed,
\begin{subequations}\label{eq:end_constraints}
	\begin{align}
		\br(1)-\br(0)&=\be_3\, ,\label{eq:end_to_end_position}\\
		\bd_1(1) - \bd_1(0)&=\bzero\, ,\label{eq:end_to_end_d_1}
	\end{align}
\end{subequations}
i.e., we fix the end-to-end position vector, as well as the relative rotation of $\bd_1$ between the two ends.
Using \eqref{eq:centerline_kinematics}, constraints \eqref{eq:end_constraints} can be given the following integral forms,
\begin{subequations}
	\begin{align}
		\int_0^1\!\partial_s\br ds = \int_0^1\! v_3\bd_3\, ds= \be_3\, ,\label{eq:boundary_condition_position}\\
		\int_0^1\!\partial_s\bd_1\, ds  = \int_0^1(u_3\bd_2 - u_2\bd_3)\,ds = \bzero\, .\label{eq:boundary_condition_director}
	\end{align}
\end{subequations}
We restrict our interest to the cases where the centerline deforms in the $\{\be_2,\be_3\}$ plane only, meaning that the deformed centerline can be accorded the following representation,
\begin{align}
	\br = x_2\be_2 + x_3\be_3\, .\label{eq:centerline_planar}
\end{align}
In line with this assumption of planar deformations, we further restrict the set of possible deformations with the following conditions on the strains,
\begin{align}
	u_2=u_3=0\, ,\label{eq:strains_assumptions}
\end{align}
which makes \eqref{eq:u_3} satisfied for any $\alpha$.
Boundary condition \eqref{eq:boundary_condition_director} is identically satisfied under this assumption, as the only non-zero component left of $\bu$ is $u_1$.
Equations \eqref{eq:strains_assumptions}, \eqref{eq:u2}, \eqref{eq:u_3}, and \eqref{eq:alpha0} together imply the following for non-zero $u_1$,
\begin{align}
	\alpha(s) = \alpha_0(s) = 0\, ,\label{eq:alphas}
\end{align}

Next, we parameterize $\bd_3$ and $\bd_2$ using an angle $\vartheta$ as follows,
\begin{subequations}\label{eq:directors_parametrisation}
	\begin{align}
		\bd_2 &= \cos\vartheta\be_2 - \sin\vartheta\be_3\, ,\label{eq:d2_parametrisation}\\
		\bd_3 &= \sin\vartheta\be_2 + \cos\vartheta\be_3\, .\label{eq:d3_parametrisation}
	\end{align}
\end{subequations}
It is then easy to deduce using the above and \eqref{eq:centerline_kinematics}$_2$ the following parametrization for the only non-zero component of the strain vector,
\begin{equation}
	u_1=\vartheta'\, ,\label{eq:strain_parametrization}
\end{equation}
where a prime $'$ denotes differentiation with respect to $s$.

Using \eqref{eq:alphas} and \eqref{eq:strain_parametrization}, the functions $f_1$ and $f_3$ in \eqref{eq:energy_densities} reduce to,
\begin{equation}
	f_1 = v_3^2 + \frac{\nu^2}{v_3^2}\quad\text{and}\quad	f_3 = \frac{4}{3}\left(\frac{\vartheta'}{v_3}\right)^2\, ,\label{eq:energy_densities_rectangular}
\end{equation}
where
\begin{equation}
	\label{eq:nu_definition}
	\nu := \sqrt{\frac{S_0+1}{S+1}}
\end{equation}
is the effective activation parameter.\footnote{Note that $\nu$ defined here is the reciprocal of $\mu$ defined in \cite{singh:model}. The convenience of such a choice will become clear from the appearance of the bifurcation diagrams in Fig.~\ref{fig:bifurcation_diagrams}.} 
With this definition of $\nu$, the centerline of the ribbon would be expected to expand when $\nu>1$.
The function $f_1$, representing the stretching energy density in the ribbon, attains its minimum at $v^*_3 = \sqrt{\nu}$ (see also Appendix~\ref{sec:stretching}).
The coupling between bending and stretching in the ribbon under consideration is clearly delineated in the expression for $f_3$, where the stretch $v_3$ appears in the denominator.
This functional dependence of $f_3$ on $v_3$ effectively reduces the bending rigidity of the ribbon along the length in a variable fashion.

The energy functional \eqref{eq:energy} augmented with the constraint \eqref{eq:boundary_condition_position} can then be written as follows in the unknown functions $\vartheta(s)$ and $v_3(s)$,
\begin{align}
	\widetilde{F}[\vartheta,v_3]:=\int_0^1\left(f_1 h  + f_3 h^3 + \Lambda_2v_3\sin\vartheta + \Lambda_3( v_3\cos\vartheta- 1)\right)ds \, ,\label{eq:simplified_energy}
\end{align}
where $\Lambda_i={\bm\Lambda}\cdot\be_i$, $i\in\{2,3\}$, are the Cartesian components of a constant vector multiplier $\bm\Lambda$ employed to enforce \eqref{eq:boundary_condition_position}. 

Using standard variational procedure, we compute the variation in $\widetilde{F}[\vartheta,v_3]$ due to $\vartheta\rightarrow\vartheta+\delta\vartheta$ as follows,
\begin{equation}
	\begin{split}
	\delta\widetilde{F}(\vartheta,v_3)[\delta\vartheta] &= \left(\frac{8h^3}{3v_3^2}\vartheta'\delta\vartheta\right)\bigg\vert_0^1 \\&- \int_0^1\left\{\left(\frac{8h^3}{3v_3^2}\vartheta'\right)' - \Lambda_2 v_3\cos\vartheta + \Lambda_3 v_3 \sin\vartheta\right\}\delta\vartheta \,ds\, ,\label{eq:theta_variation}
	\end{split}
\end{equation}
delivering the Euler-Lagrange (EL) equation for $\vartheta$ along with the corresponding boundary conditions,

\begin{subequations}
	\begin{align}
		\left(\frac{8h^3}{3v_3^2}\vartheta'\right)' - \Lambda_2 v_3\cos\vartheta + \Lambda_3 v_3 \sin\vartheta &= 0\, ,\label{eq:EL_theta}\\
		\left(\frac{8h^3}{3v_3^2}\vartheta'\right)_{s=0} = \left(\frac{8h^3}{3v_3^2}\vartheta'\right)_{s=1} &= 0\, . \label{eq:bc_u1}
	\end{align}
\end{subequations}
Similarly, the EL equation corresponding to $v_3$ is given by a straightforward partial derivative in $v_3$ of the integrand of \eqref{eq:simplified_energy},
\begin{align}
	\frac{8h^3\vartheta'^2}{3 v_3^3} + 2 h\left(\frac{\nu^2}{ v_3^3} - v_3\right) - \Lambda_2\sin\vartheta - \Lambda_3\cos\vartheta = 0\, .\label{eq:v3}
\end{align}
There are no boundary conditions associated with $v_3$.

\subsection{Equilibrium equations}\label{sec:governing_equations}
Here we obtain a complete system of first order ODEs representing our equilibrium  problem. 
The state space of our system is described by the vector $\{\vartheta,u_1,v_3,x_2,x_3,\Lambda_2,\Lambda_3\}$, with $7$ fields.
The evolution of $\vartheta,u_1,x_2,x_3$ with $s$ can be easily obtained using \eqref{eq:strain_parametrization}, \eqref{eq:EL_theta}, \eqref{eq:centerline_kinematics}$_1$, \eqref{eq:centerline_planar}, and \eqref{eq:d3_parametrisation}.
The two multipliers $\Lambda_2$ and $\Lambda_3$ are constant, so their evolution is  trivial.
The only field whose evolution  needs to be determined is $v_3$, as the corresponding EL equation \eqref{eq:v3} is algebraic in $v_3$.
To that end, we differentiate \eqref{eq:v3} and use it to resolve for $u'_1$ and $v_3'$ using \eqref{eq:EL_theta} (and \eqref{eq:strain_parametrization} to eliminate $\vartheta'$).
As a result, we arrive at the following set of first order ODEs,
\begin{subequations}\label{eq:full_system}
	\begin{align}
		\vartheta' & = u_1\, ,\label{eq:simple_dtheta}\\
		u'_1&=\frac{-3v_3^3(\Lambda_2\cos\vartheta - \Lambda_3\sin\vartheta)(9\nu^2 + 4 h^2 u_1^2 + 3 v_3^4)}{8h^3\left[ 4h^2 u^2_1 - 3(3\nu^2+ v_3^4)\right]}\, ,\label{eq:simple_du1}\\
		v_3'&=\frac{-3u_1 v_3^4(\Lambda_2\cos\vartheta - \Lambda_3\sin\vartheta) }{8h^3 u_1^2 - 6h(3\nu^2 + v_3^4)}\, .\label{eq:simple_dv3}\\
		x_2'&=v_3\sin\vartheta\, ,\label{eq:simple_dx2}\\
		x_3' &= v_3\cos\vartheta\, .\label{eq:simple_dx3}\\
		\Lambda_2' &= 0\, ,\label{eq:simple_dLambda2}\\
		\Lambda_3'&=0\, .\label{eq:simple_dLambda3}
	\end{align}
\end{subequations}
These 7 ODEs need to be complemented with $7$ boundary conditions.
Two of these are given by \eqref{eq:bc_u1}, and the rest are the following 5,
\begin{equation}
	\begin{split}
	x_2(0) = 0\, ,\quad x_2(1) = 0\, ,\quad x_3(0)=0\, ,\quad x_3(1) = 1\, ,\\ 2h\left(\frac{\nu^2 }{v_3(0)^3} - v_3(0)\right) - \Lambda_2(0) \sin\vartheta(0) - \Lambda_3(0) \cos\vartheta(0)=0\, ,\label{eq:bc_additional}
	\end{split}
\end{equation}
where the last condition ensures that the boundary condition on $v_3(0)$ is consistent with the parent algebraic equation \eqref{eq:v3} from which \eqref{eq:simple_dv3} was derived.
Our system is now complete, and can be viewed as a two point boundary value problem with $\nu$ as a control parameter. 

Physical intuition suggests that for $\nu=1$, that is, when $S=S_0$, no spontaneous deformation of the ribbon should take place, as the material is not activated. In the absence of deformation, the ribbon remains in the straight configuration, where $\vartheta\equiv0$, $u_1\equiv0$, and $v_3\equiv1$, which solve \eqref{eq:full_system} with $\Lambda_2=\Lambda_3=0$ and $x_2=0$, $x_3=s$. It is indeed a simple matter to show that the configuration
\begin{equation}
	\begin{split}
	\vartheta=0,\quad u_1=0,\quad v_3=1,\quad\Lambda_2=0,\quad \Lambda_3=2h(\nu^2-1),\\ x_2=0,\quad x_3=s\, ,\label{eq:trivial_solution}
	\end{split}
\end{equation}
is a solution of \eqref{eq:full_system} for all $h>0$ and $\nu>0$. This is the \emph{trivial} solution of our equilibrium problem and we would like to obtain solutions to \eqref{eq:full_system} \emph{emanating} from it. 

The system of governing equations \eqref{eq:full_system} is invariant under the transformations $\vartheta\mapsto-\vartheta,\,u_1\mapsto-u_1,\,v_3\mapsto v_3,\,\Lambda_2\mapsto -\Lambda_2,\,\Lambda_3\mapsto \Lambda_3,\, x_2\mapsto -x_2\,\,x_3\mapsto x_3$. Thus, for every equilibrium configuration $\left(x_3(s),x_2(s)\right)$, its mirror image $\left(x_3(s),-x_2(s)\right)$ is also a valid equilibrium configuration. All equilibrium configurations come in mirror-symmetric pairs.

\subsection{Bifurcation analysis}\label{sec:bifurcation_analysis}
Consider the following solution to \eqref{eq:full_system} which differs from the trivial solution \eqref{eq:trivial_solution} by an infinitesimal amount,
\begin{equation}
\begin{split}
	\vartheta=\varepsilon\tvartheta,\ u_1=\varepsilon\tu_1,\ v_3=1+\varepsilon\tv_3,\ \Lambda_3=2h(\nu^2-1)+\varepsilon\tLambda_3,\ \Lambda_2=\varepsilon\tLambda_2,\\ x_2=\varepsilon\tx_2,\ x_3=s+\varepsilon\tx_3\, ,\label{eq:perturbed_solution}
\end{split}
\end{equation}
where $\varepsilon$ is a small non-zero perturbation parameter.
Using \eqref{eq:bc_u1} and \eqref{eq:bc_additional}, we deduce the following boundary conditions on the perturbations,
\begin{align}
	\tvartheta'(0)=\tvartheta'(1)=0,\quad \tx_2(0)=\tx_2(1)=0,\quad\tx_3(0) = \tx_3(1) = 0\, ,\label{eq:perturbations_bc}
\end{align}
Substituting \eqref{eq:perturbations_bc} into \eqref{eq:full_system}, and equating equal powers of $\varepsilon$, we obtain,
\begin{subequations}\label{eq:perturbed_system}
	\begin{align}
		\tvartheta' &= \tu_1\, ,\label{eq:ttheta}\\
		\tu'_1 &=\frac{3}{8h^3}\left[\tLambda_2-2h(\nu^2 - 1)\tvartheta\right]\, ,\label{eq:tu1}\\
		\tv'_3&=0\, ,\label{eq:tv3}\\
		\tx'_2&=\tvartheta\, ,\label{eq:tx2}\\
		\tx'_3&=\tv_3\, ,\label{eq:tx3}\\
		\tLambda_2'&=0\, ,\label{eq:tLambda2}\\
		\tLambda'_3&=-2h(3\nu^2 + 1)\tv_3\, .\label{eq:tLambda3}
	\end{align}
\end{subequations}
The boundary conditions on $\tx_3$ in \eqref{eq:perturbations_bc} imply that $\tv_3$ vanishes (from \eqref{eq:tv3} and \eqref{eq:tx3}), and so does $\tLambda_3$ (from \eqref{eq:tLambda3}).
Using \eqref{eq:ttheta} and \eqref{eq:tu1} we arrive at the following second-order linear ODE for $\tvartheta$,
\begin{align}
	\tvartheta'' + \frac{3}{4h^2}\left(\nu^2-1\right)\tvartheta = \frac{3\tLambda_2}{8h^3}\, .\label{eq:perturbation_ODE}
\end{align}
For $\nu <1$, the above ODE has no solutions that satisfy the boundary conditions on $\tvartheta'$  stated in \eqref{eq:perturbations_bc}.
The implication being that all such configurations are likely to be locally \emph{stable}, as no equilibrium solution emanates from it. This was to be expected, as $\nu<1$ implies that $S>S_0$, and so the activated configuration is more ordered than the reference configuration, entailing an extension of the ribbon along $\bd_1$ (see Fig.~\ref{fig:rectangular_schematics}), and by the surface inextensibility constraint, a contraction along $\bd_3$, which is incompatible with boundary condition \eqref{eq:end_to_end_position}.\footnote{Here, we effectively regard $\nu$ as a \emph{continuation} parameter, which drives new equilibrium solutions out of the trivial one.}

On the other hand, for $\nu\ge 1$, which is the only case that interests us here, equation \eqref{eq:perturbation_ODE} admits non-trivial solutions consistent with the boundary conditions in \eqref{eq:perturbations_bc} when $\nu=\nu_n$, where
\begin{equation}
	\nu_n := \sqrt{1+\frac{4n^2\pi^2 h^2}{3}}\quad\text{with}\quad n\in\mathbb{N}\, .\label{eq:bifurcation_points}
\end{equation}
Consequently, $\tx_2$ is the \emph{jet} (eigenfunction) generated by $\tx_2 = \sin(n\pi s)$.
Expression \eqref{eq:bifurcation_points} indicates that for a given branch, identified by an integer value of $n$, the separation between two adjacent values of $\nu_n$ must increase with increasing value of $h$.
In the next section, we quantify this effect, illuminating the interplay of two \emph{tendencies} in the mechanical behavior of an activated ribbon, which for brevity we call ``membrane-like'' and ``plate-like''. The former tendency is naively expected to prevail for small values of $h$, where the stretching energy is dominant and the bending energy would be regarded as a perturbation; the latter, on the contrary, is expected to play a role for larger values of $h$. We shall see how   the  activation parameter can be dysfunctional towards  such naive expectations.

\section{Activated Equilibria}\label{sec:results}
The governing set of equations \eqref{eq:full_system}, accompanied by the boundary conditions \eqref{eq:bc_u1} and \eqref{eq:bc_additional}, were solved numerically using a parameter continuation method \cite{doede:numerical}, implemented in \texttt{AUTO-07P} \cite{auto07p}.
The bifurcation parameter was chosen to be the effective activation parameter $\nu$ in \eqref{eq:nu_definition}, which was varied from $1.0$ to $2.0$, and the trivial solution \eqref{eq:trivial_solution} was given as the base solution from which the continuation was carried out.

We resolve the degeneracy induced by mirror symmetry by selecting a single member in each symmetric solution pair. Each equilibrium profile shown below should ideally be accompanied by its mirror image.

\begin{figure}[h]
	\centering
	\begin{subfigure}[t]{0.48\linewidth}
		\centering
		\includegraphics[width=\linewidth]{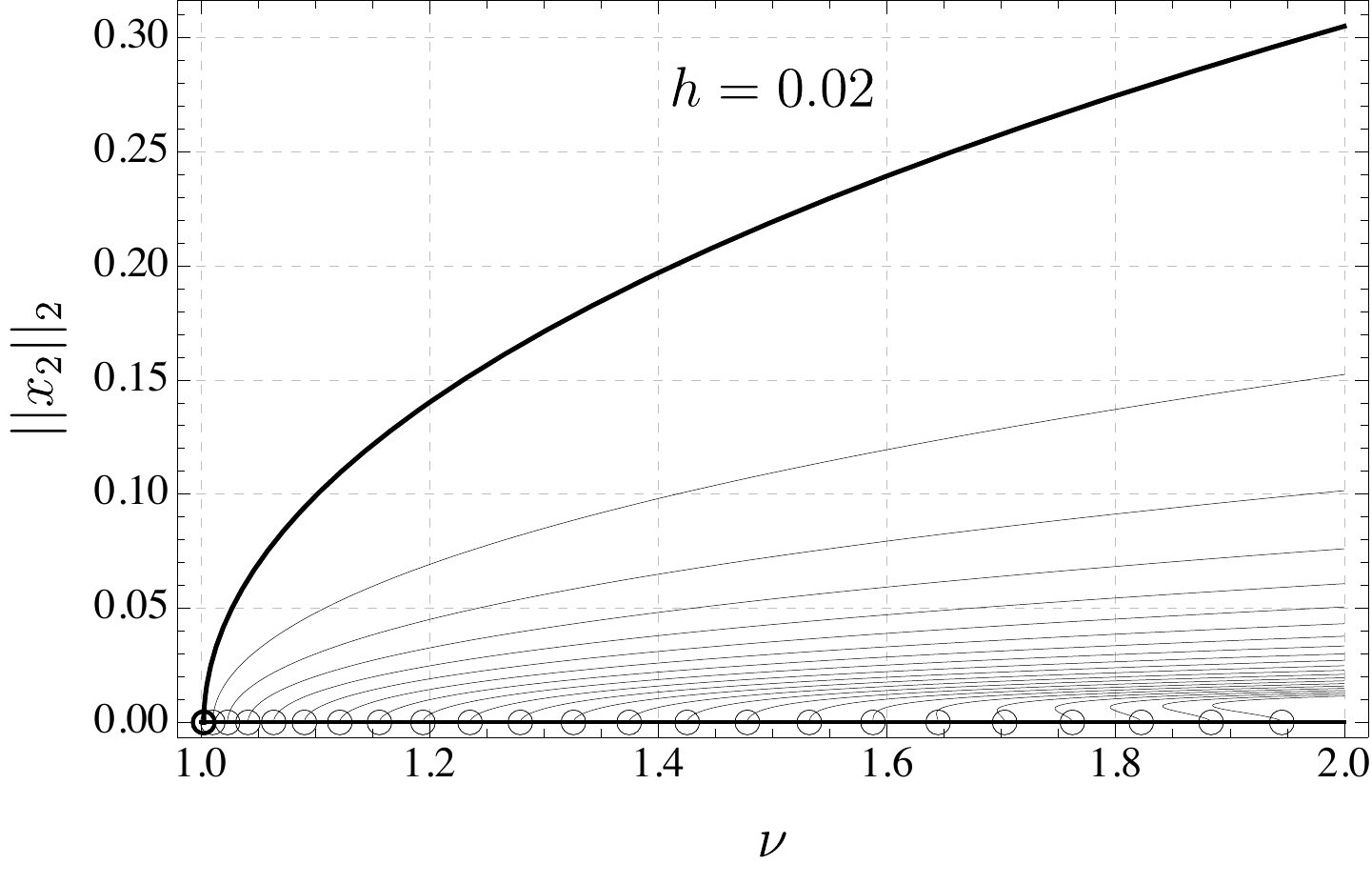}
		\caption{}\label{fig:bifurcation_diagrams_a}
	\end{subfigure}\quad
	\begin{subfigure}[t]{0.48\linewidth}
		\centering
		\includegraphics[width=\linewidth]{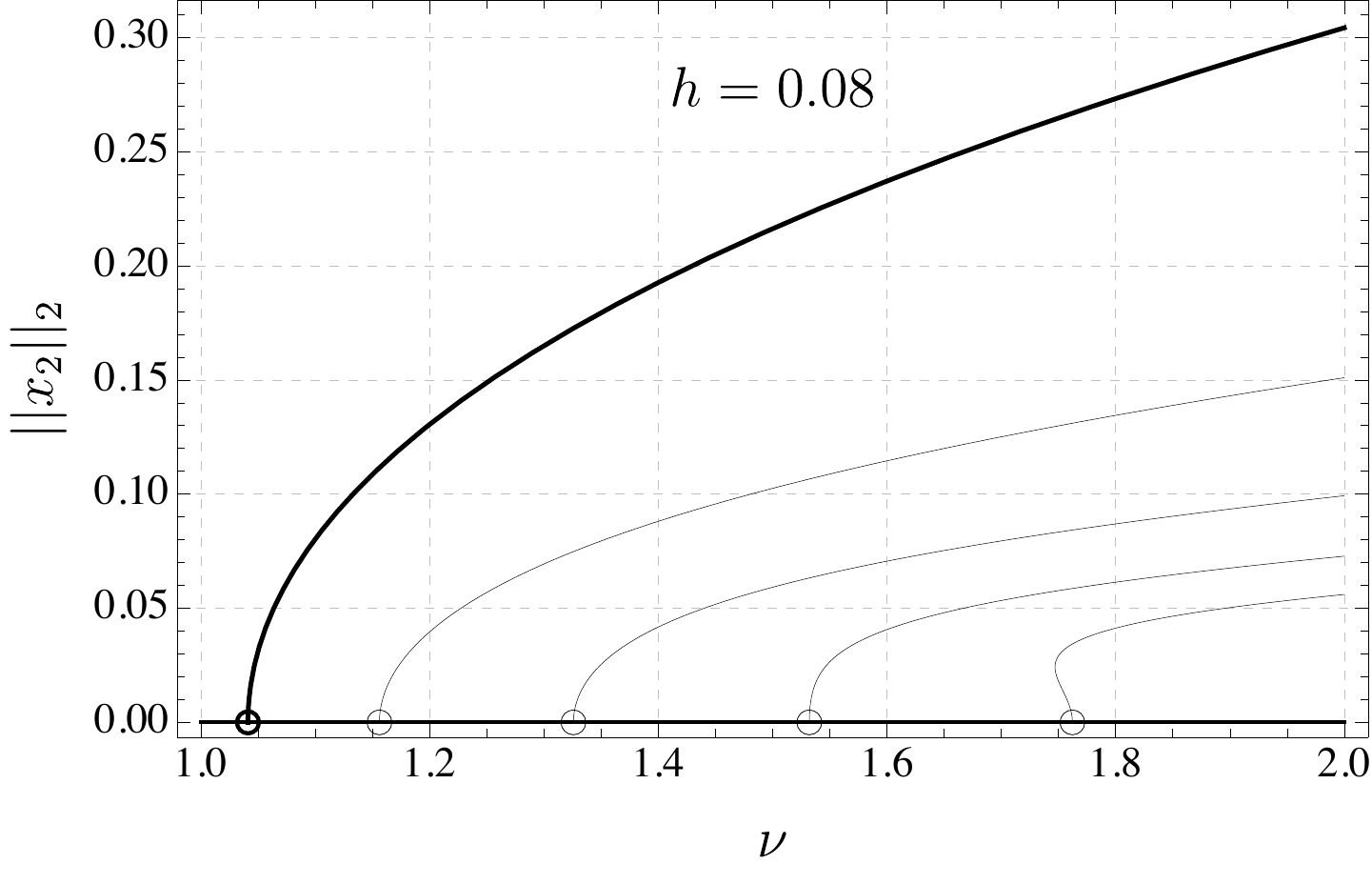}
		\caption{}\label{fig:bifurcation_diagrams_b}
	\end{subfigure}\quad
	\begin{subfigure}[t]{0.48\linewidth}
		\centering
		\includegraphics[width=\linewidth]{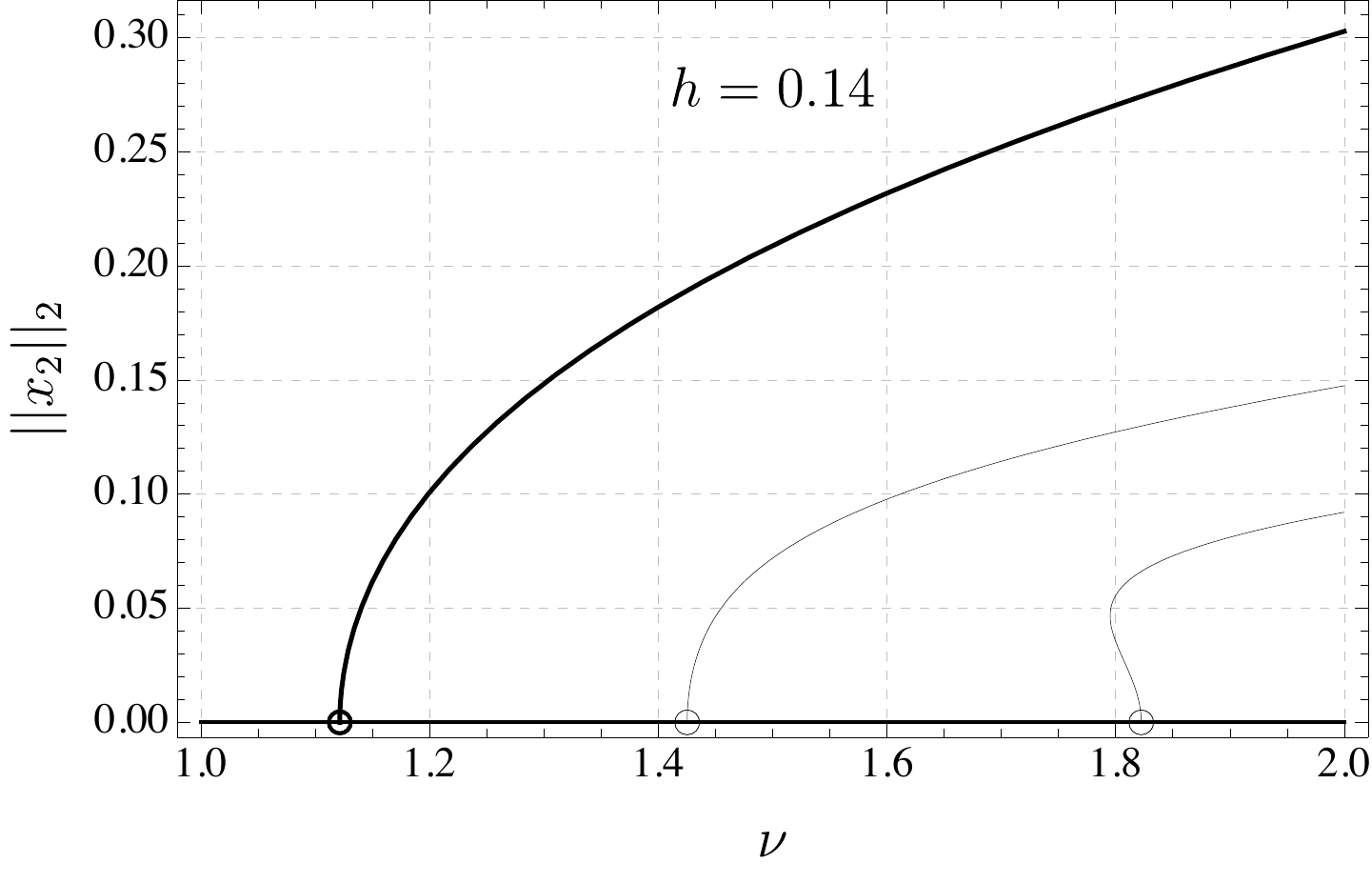}
		\caption{}\label{fig:bifurcation_diagrams_c}
	\end{subfigure}\quad
	\begin{subfigure}[t]{0.48\linewidth}
		\centering
		\includegraphics[width=\linewidth]{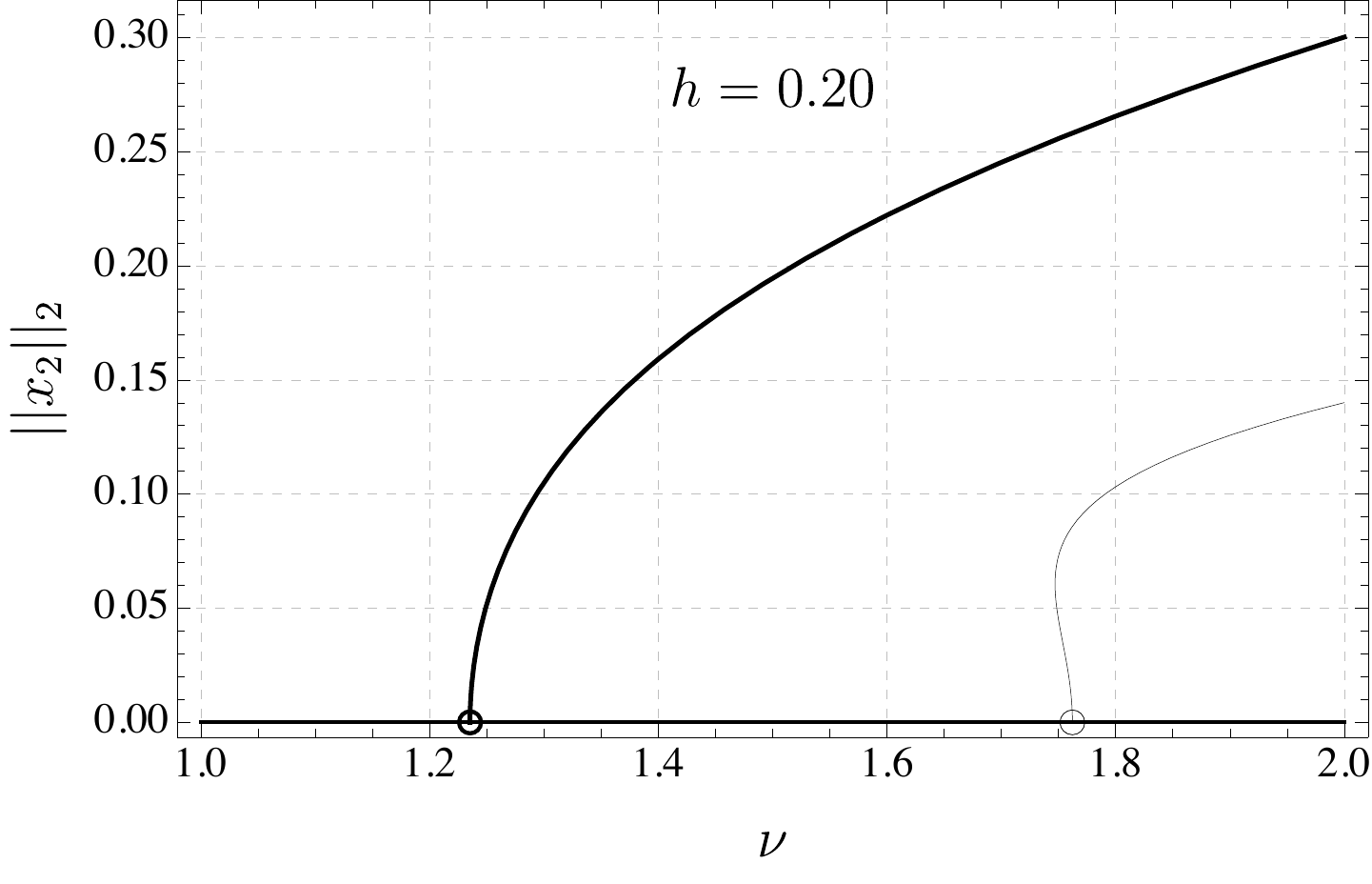}
		\caption{}\label{fig:bifurcation_diagrams_d}
	\end{subfigure}\quad
	\caption{Plots in panels (\ref{fig:bifurcation_diagrams_a}) to (\ref{fig:bifurcation_diagrams_d}) show bifurcation diagrams of the ribbon with increasing thickness. The curves are generated numerically using \texttt{AUTO-07P}, a continuation and bifurcation software for ODEs, while the circles denote the analytical predictions of the bifurcation points made using equation \eqref{eq:bifurcation_points}. The thick curve in all the plots corresponds to the first mode of deformation with one crest (see Fig.~\ref{fig:configurations_a}). The spacing between bifurcation points increases with thickness, as predicted by equation \eqref{eq:bifurcation_points}.}
	\label{fig:bifurcation_diagrams}
\end{figure} 
Our main objective behind performing the forthcoming simulations is to highlight the interplay between bending and stretching of the ribbon.
As we saw in the expressions for the bending and stretching energies in \eqref{eq:energy_densities_rectangular}, the two mechanisms of deformation are coupled via the stretch $v_3$ appearing in the denominator in the expression for $f_3$.
The following plots reveal and quantify this non-trivial coupling between the two competing modes of deformation.

We begin by tracking the evolution of the bifurcation diagrams of the ribbon as its thickness is increased.
The bifurcation plots, i.e. the $L_2$ norm of the function $x_2(s)$ vs $\nu$, generated from the numerical simulations for increasing values of the thickness $h$ are shown in Fig.~\ref{fig:bifurcation_diagrams}.
The bifurcation points $(\nu_n,0)$ as predicted by the analytical expression \eqref{eq:bifurcation_points} are shown in the plots using circles, and are in good agreement with the numerical solutions.
Figures~\ref{fig:bifurcation_diagrams_a} to \ref{fig:bifurcation_diagrams_d} show the distances between the bifurcation points 
increasing with thickness,
confirming our assertion made at the end of Sect.~\ref{sec:bifurcation_analysis}.
This increasing gap between bifurcation points is an indication of the gradual transformation of the ribbon's behavior from membrane-like to plate-like.

\begin{figure}[h]
	\centering
	\begin{subfigure}[t]{0.48\linewidth}
		\centering
		\includegraphics[width=\linewidth]{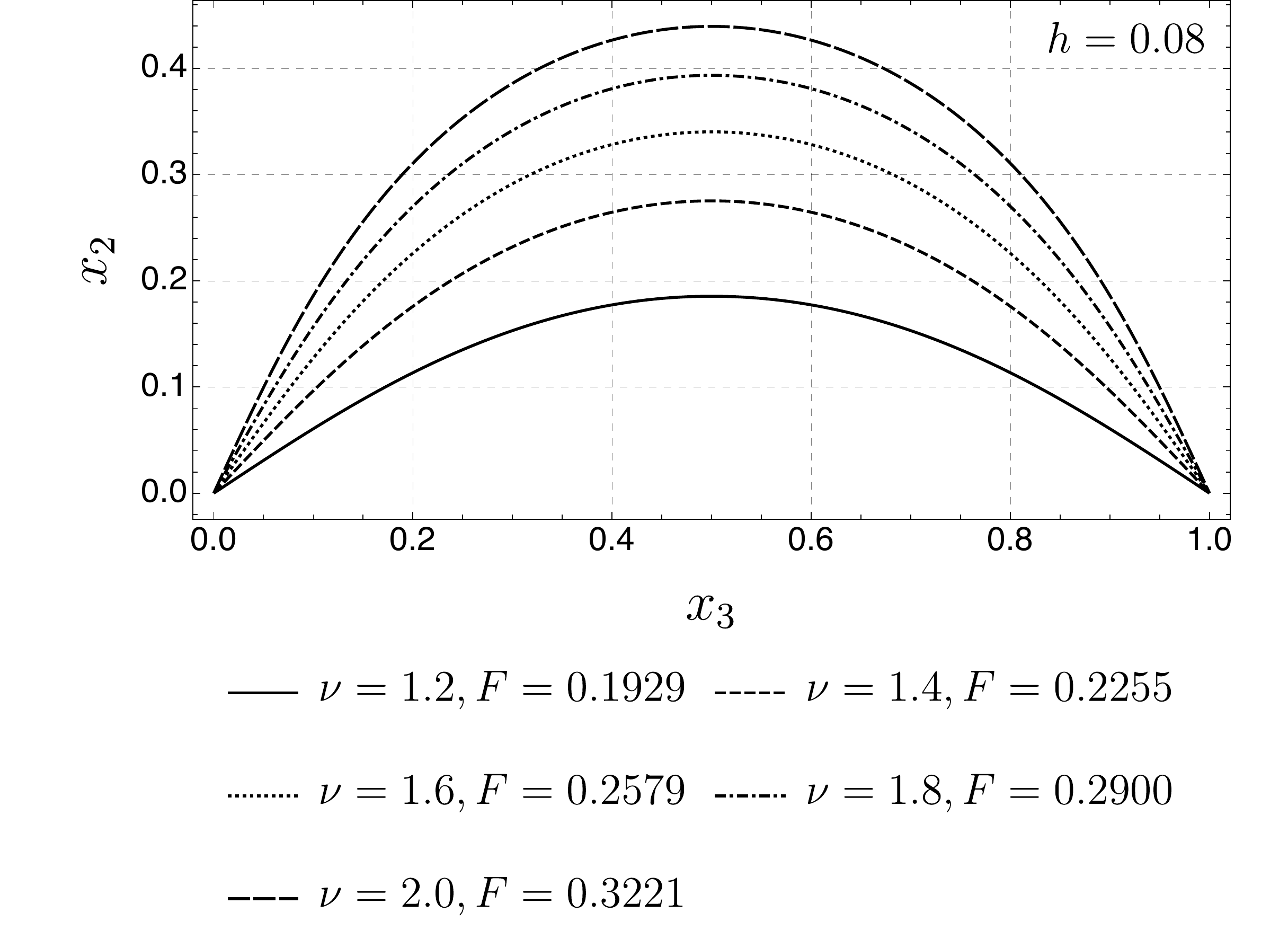}
		\caption{}\label{fig:configurations_a}
	\end{subfigure}\quad
	\begin{subfigure}[t]{0.48\linewidth}
		\centering
		\includegraphics[width=\linewidth]{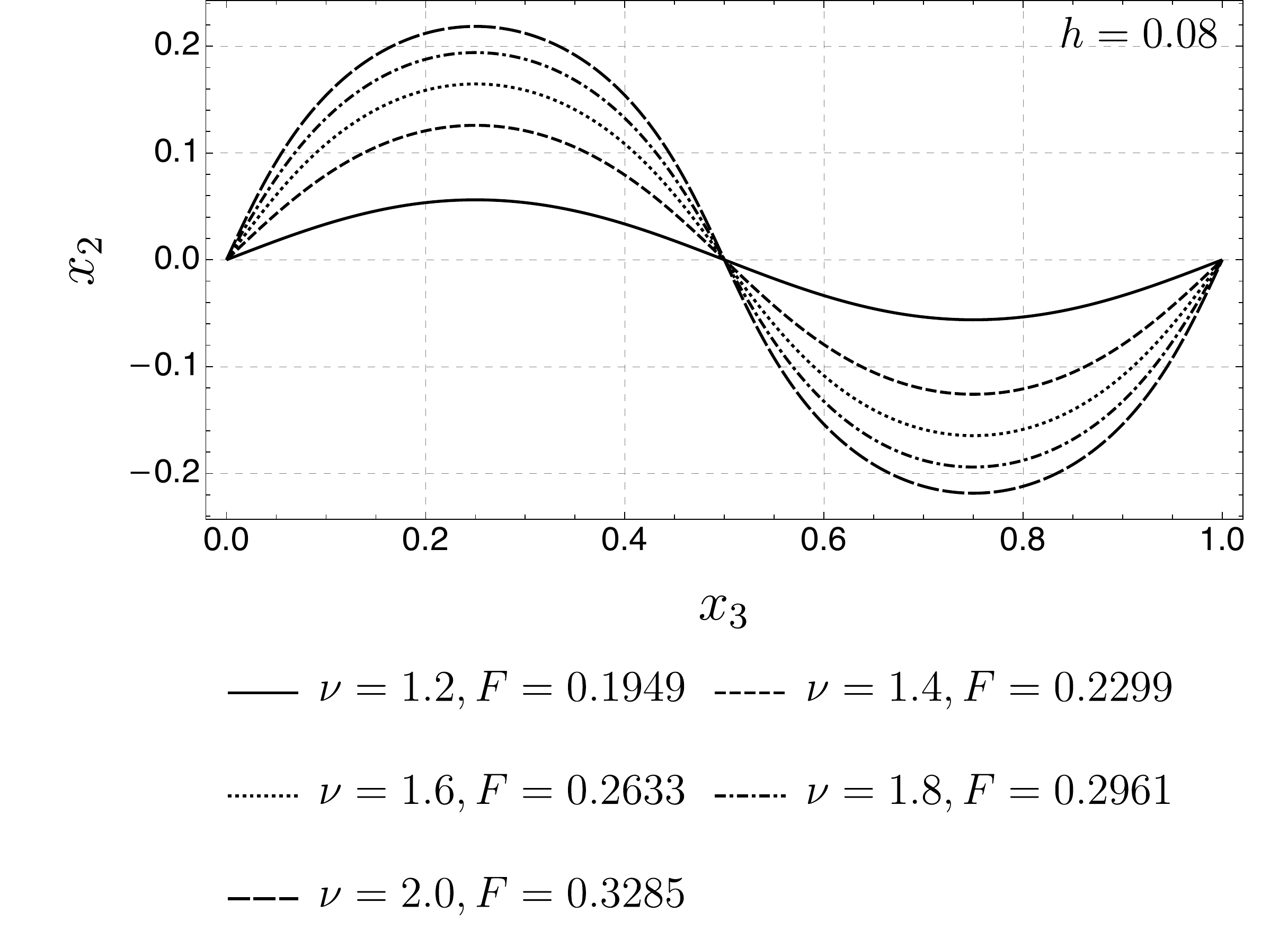}
		\caption{}\label{fig:configurations_b}
	\end{subfigure}\hfill
	\caption{Plots in panels (\ref{fig:configurations_a}) and (\ref{fig:configurations_b}), respectively, correspond to the first and the second modes of deformation of a ribbon of thickness $h=0.08$ (in units of $L$), for different values of parameter $\nu$. The total energy $F$ in \eqref{eq:energy} of each configuration is shown in the legend below each panel. For a given value of $\nu$, the energy of the first mode is very close to the energy of the second mode, however, the latter remains higher than the former in all cases, and both are lower than the energy $h(1+\nu^2)$ of the trivial solution \eqref{eq:trivial_solution}.}
	\label{fig:configurations}
\end{figure} 
\begin{figure}[h!]
	\centering
	\begin{subfigure}[t]{0.48\linewidth}
		\centering
		\includegraphics[width=\linewidth]{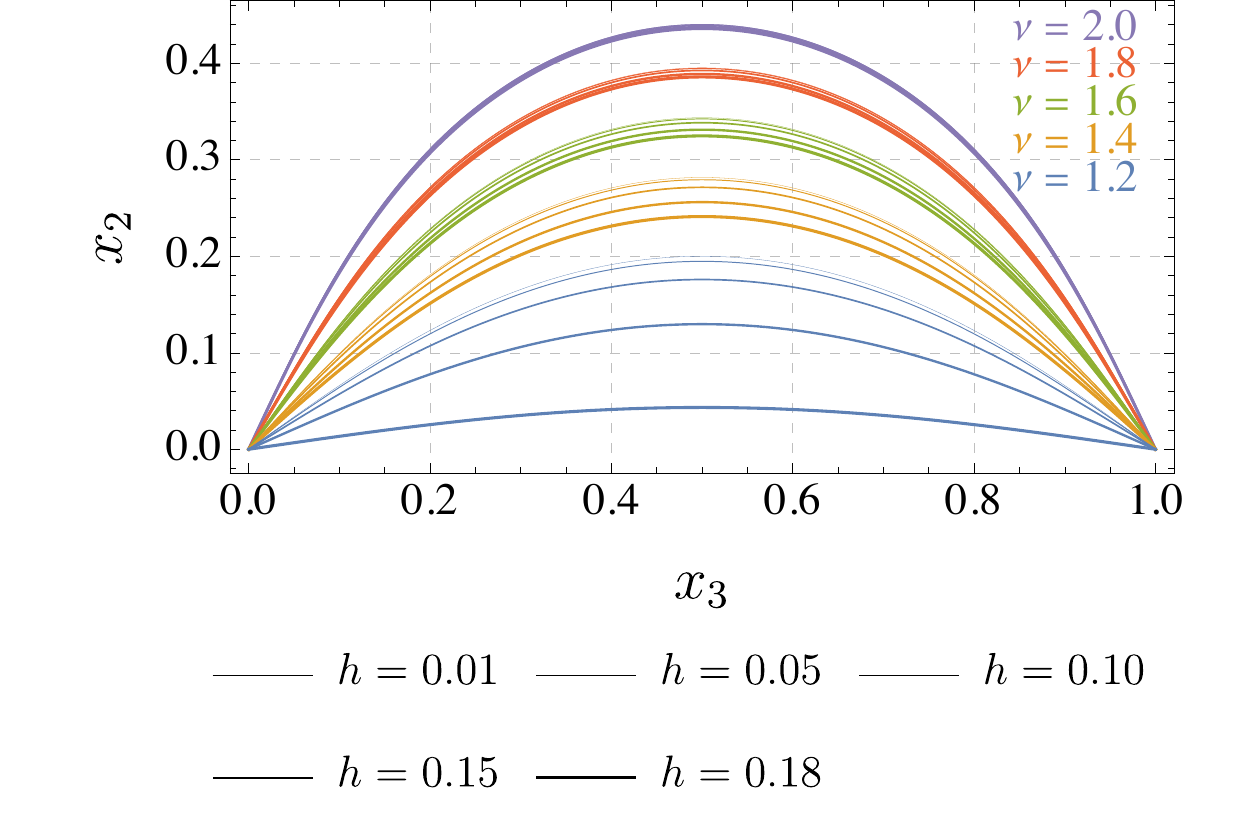}
		\caption{}\label{fig:varying_thickness_a}
	\end{subfigure}\quad
	\begin{subfigure}[t]{0.48\linewidth}
		\centering
		\includegraphics[width=\linewidth]{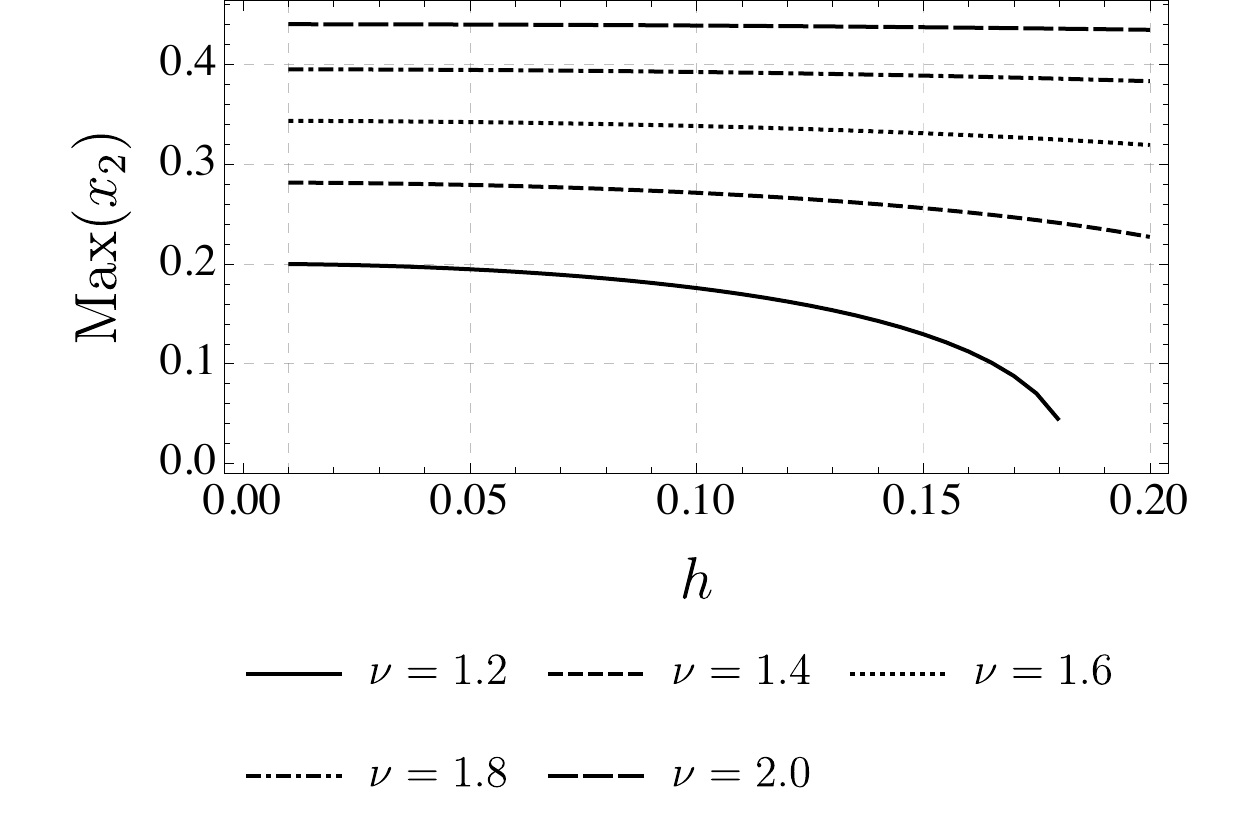}
		\caption{}\label{fig:varying_thicknesss_b}
	\end{subfigure}\hfill
	\caption{Panel (\ref{fig:varying_thickness_a}): Configurations of a thin NPN ribbon with different values of $\nu$ and thickness $h$, where each color represents a fixed value of $\nu$. In all cases, increasing thickness results in decreasing the maximum displacement of the ribbon. However, this effect appears to be more pronounced for smaller values of $\nu$. Panel (\ref{fig:varying_thicknesss_b}): Maximum displacement of the first mode of deformation of an NPN ribbon plotted against its normalized thickness for various values of $\nu$. }
	\label{fig:varying_thickness}
\end{figure} 
A series of configurations of the ribbon centerline for different values of the effective activation parameter $\nu$ are shown in Fig.~\ref{fig:configurations}.
The configurations in Fig.~\ref{fig:configurations_a} lie on the first bifurcating branch (the thickened curve) in Fig. ~\ref{fig:bifurcation_diagrams_b}, while configurations in Fig.~\ref{fig:configurations_b} lie on the second bifurcating branch (adjacent to the thickened curve) in Fig.~\ref{fig:bifurcation_diagrams_b}.
The total energy of each configuration is stated in a legend below the two plots in Fig.~\ref{fig:configurations}.
A curious observation is that for a given value of $\nu$ the energies of the first and the second mode of deformation are very close, although the latter is always higher than the former. As expected, the maximum displacement is larger for the first mode than for the second mode (a rough estimate is that the former is twice as efficient as the latter). For the first mode, the maximum displacement also dramatically increases with activation, ranging from roughly $20\%$ to above $40\%$ of the ribbon's length, as $\nu$ ranges from $1.2$ to $2$.

Next we explore the effect of thickness on the maximum displacement of the ribbon in the first mode.
We plot various configurations in Fig.~\ref{fig:varying_thickness_a} for different values of $\nu$, ranging from $1.2$ to $2.0$, and thickness $h$ ranging from $0.01$ to $0.18$ (in $L$ units).
We find that increasing the thickness of the ribbon reduces its maximum displacement for any given value of $\nu$.
However, as was perhaps to be expected, this effect is much more pronounced for smaller values of $\nu$, as can be easily observed from Fig \ref{fig:varying_thickness_a}. So much so that for $\nu=2$ the deformed shapes of the ribbon corresponding to different  thicknesses are nearly indiscernible from one another. We may say that high values of the activation parameter \emph{obliterate} the  distinction between ``membrane-like'' and ``plate-like'' behavior of the ribbon. The dependence of displacement on thickness is quantified in Fig.~\ref{fig:varying_thicknesss_b}, which shows  the maximum displacement of the ribbon centerline plotted against its thickness for several values of the effective activation parameter.

\begin{figure}[h!]
	\centering
	\begin{subfigure}[t]{0.30\linewidth}
		\centering
		\includegraphics[width=\linewidth]{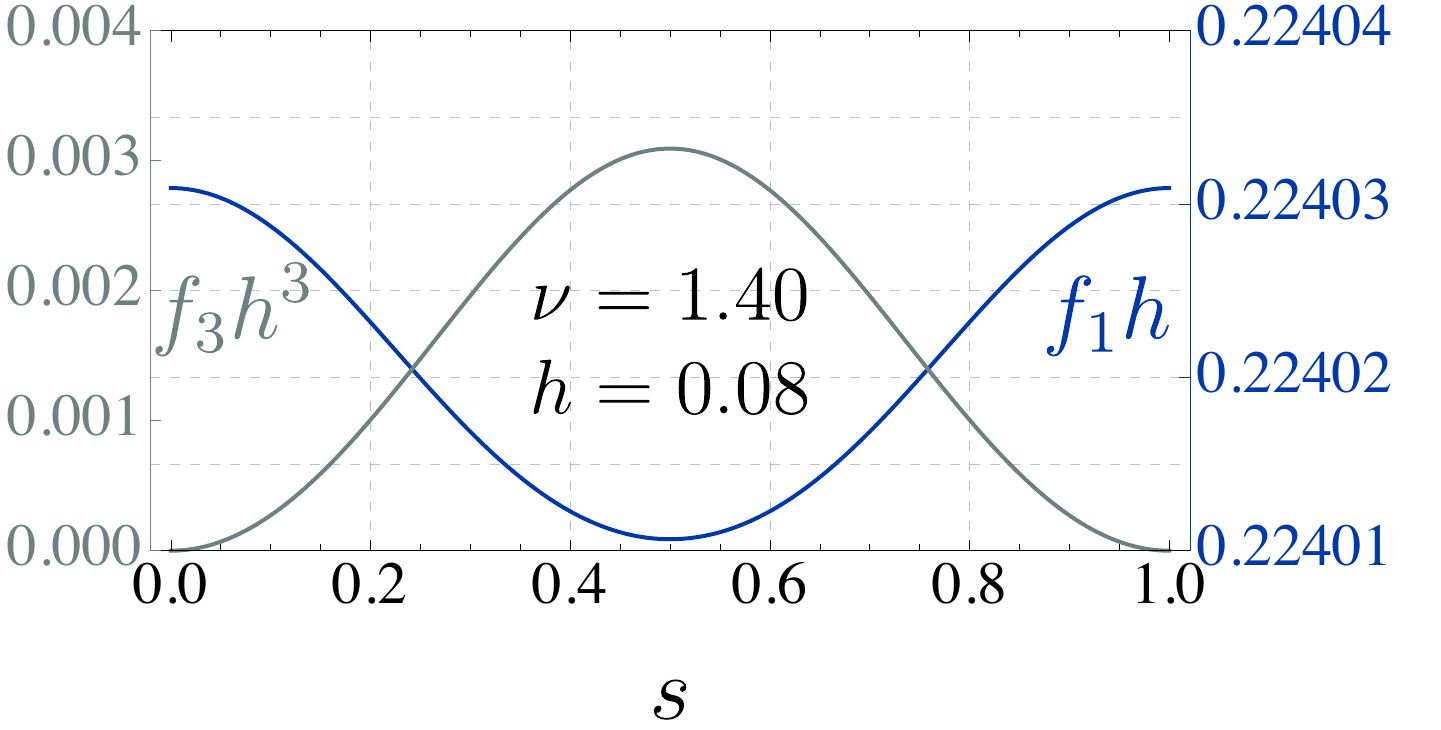}
		\caption{}\label{fig:energies_comparison_a}
	\end{subfigure}\quad
	\begin{subfigure}[t]{.30\linewidth}
		\centering
		\includegraphics[width=\textwidth]{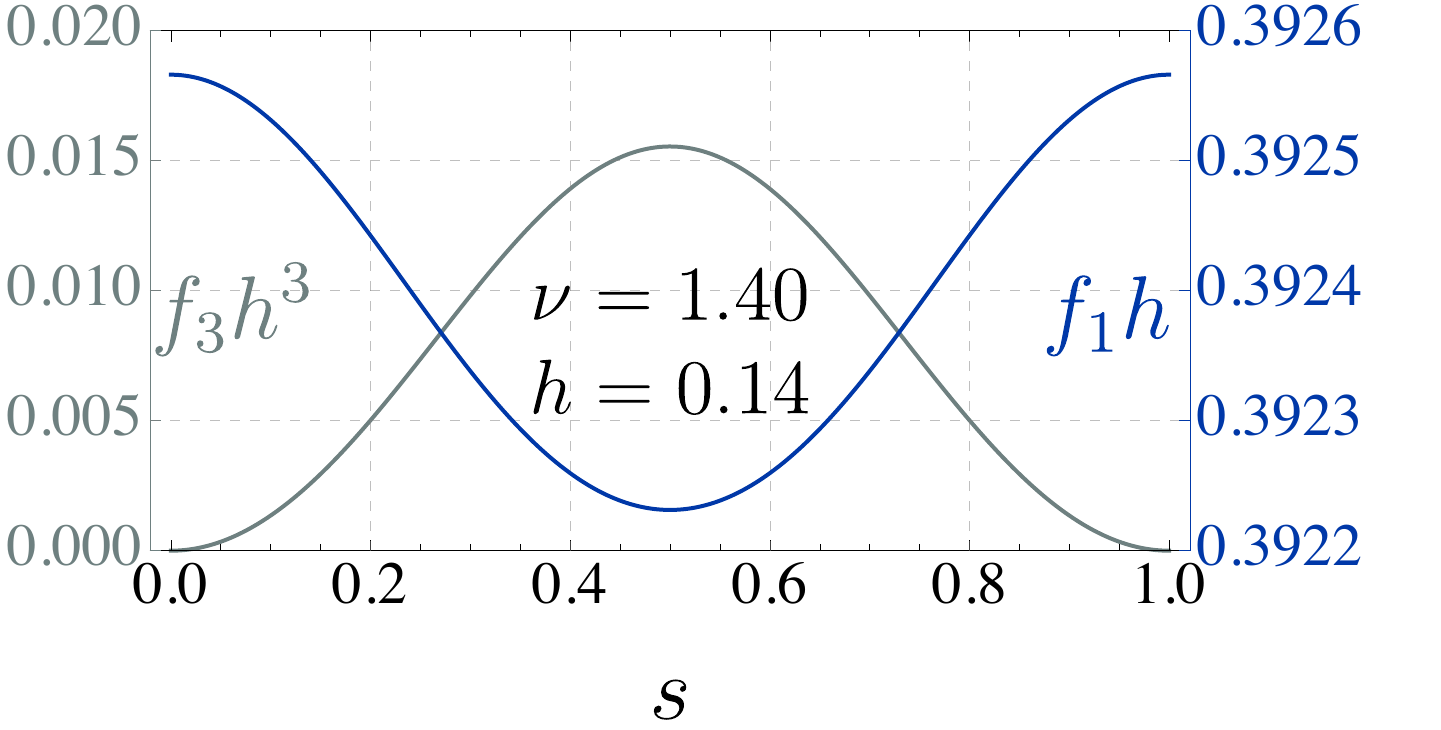}
		\caption{}\label{fig:energies_comparison_b}
	\end{subfigure}
	\begin{subfigure}[t]{.30\linewidth}
		\centering
		\includegraphics[width=\textwidth]{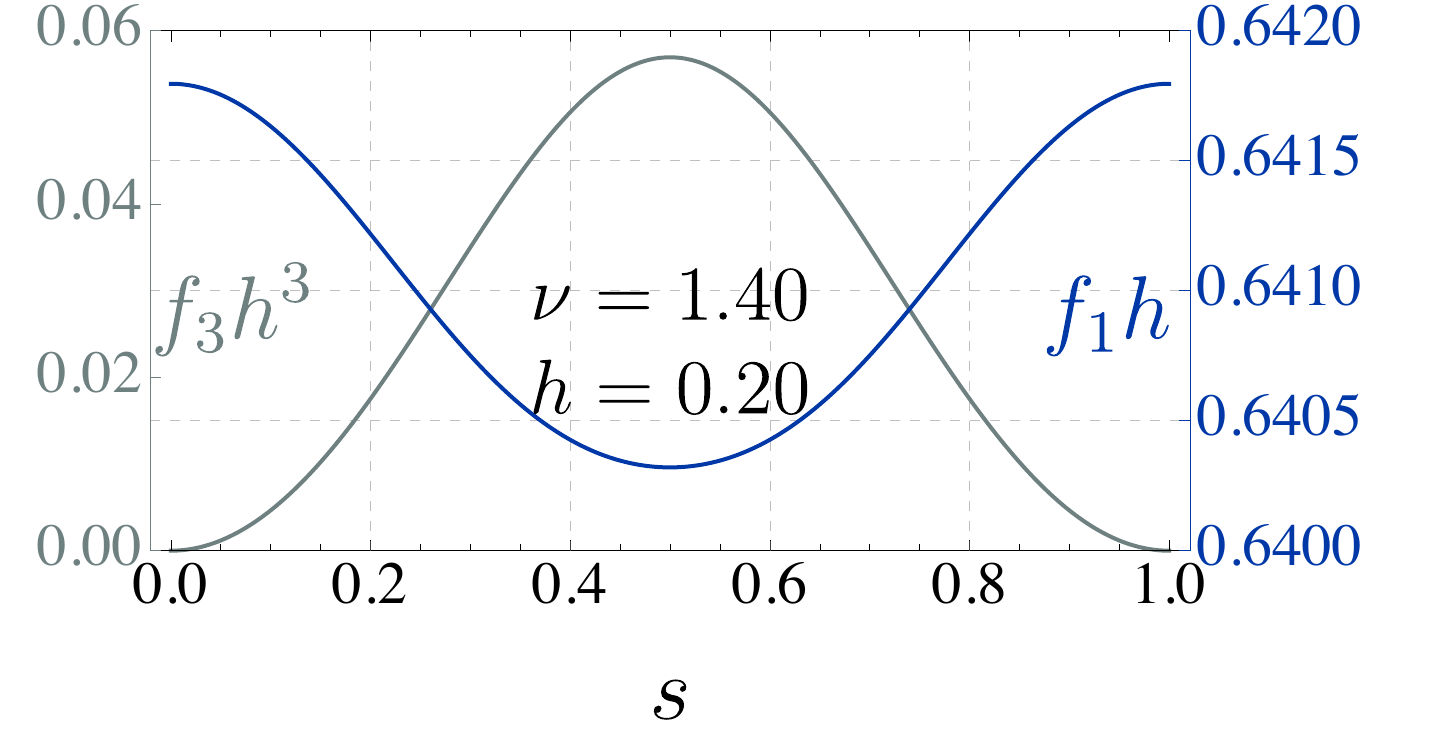}
		\caption{}\label{fig:energies_comparison_c}
	\end{subfigure}\hfill
	\begin{subfigure}[t]{0.30\linewidth}
		\centering
		\includegraphics[width=\linewidth]{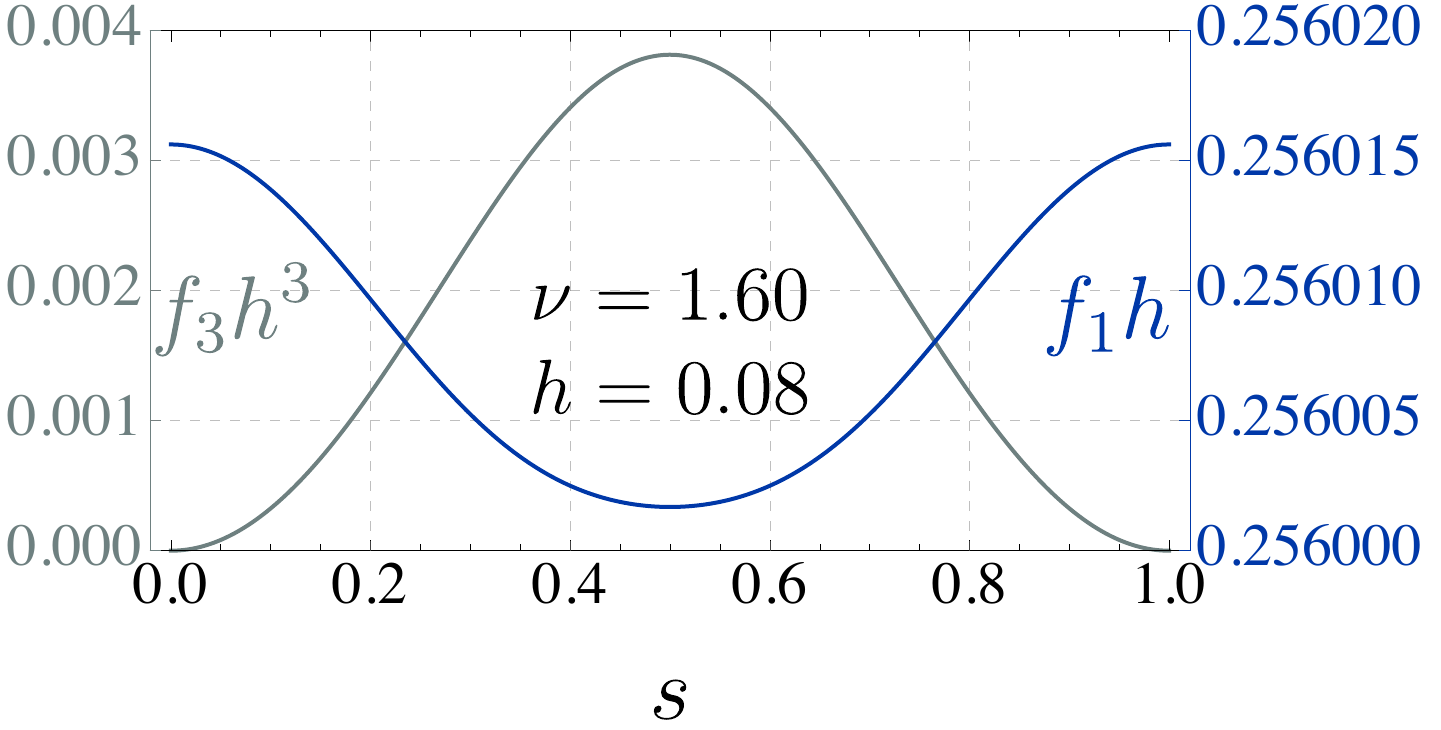}
		\caption{}\label{fig:energies_comparison_d}
	\end{subfigure}\quad
	\begin{subfigure}[t]{.30\linewidth}
		\centering
		\includegraphics[width=\textwidth]{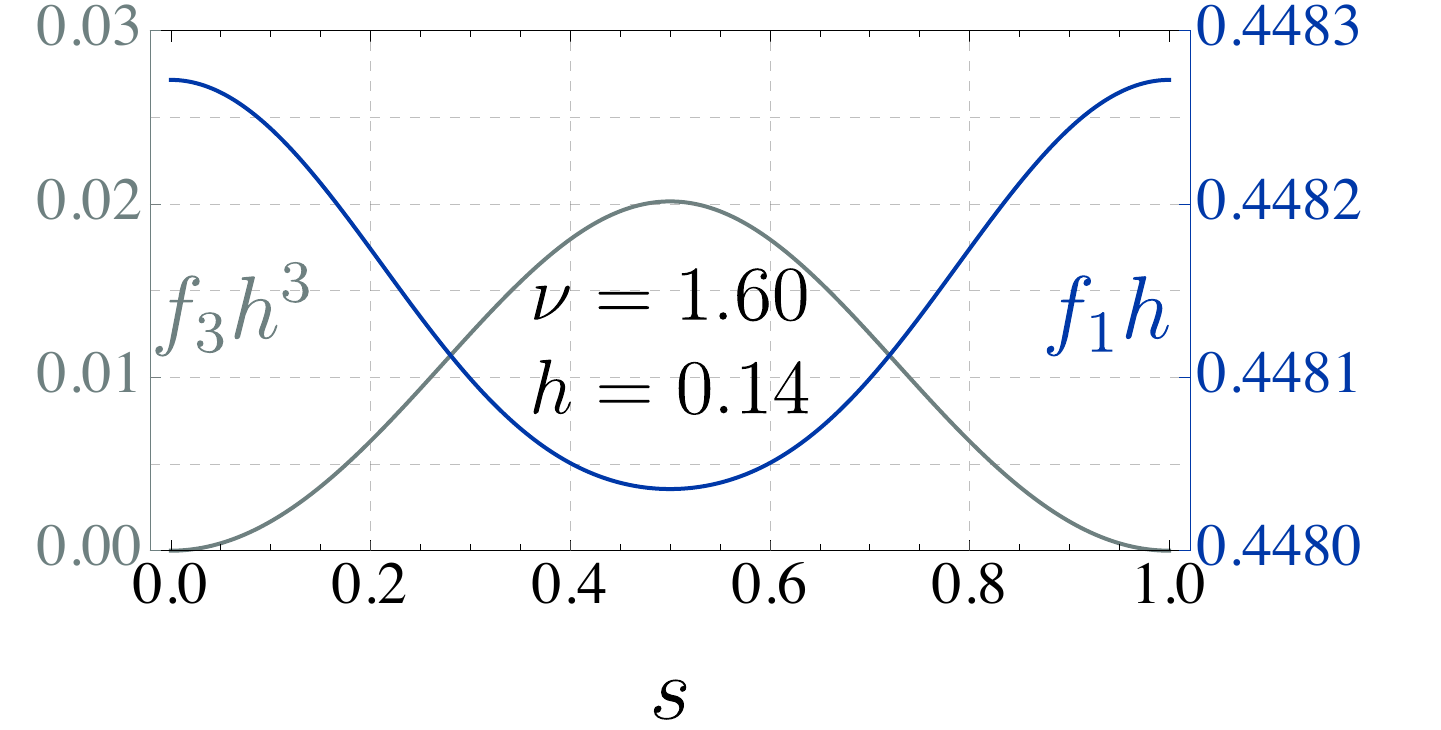}
		\caption{}\label{fig:energies_comparison_e}
	\end{subfigure}
	\begin{subfigure}[t]{.30\linewidth}
		\centering
		\includegraphics[width=\textwidth]{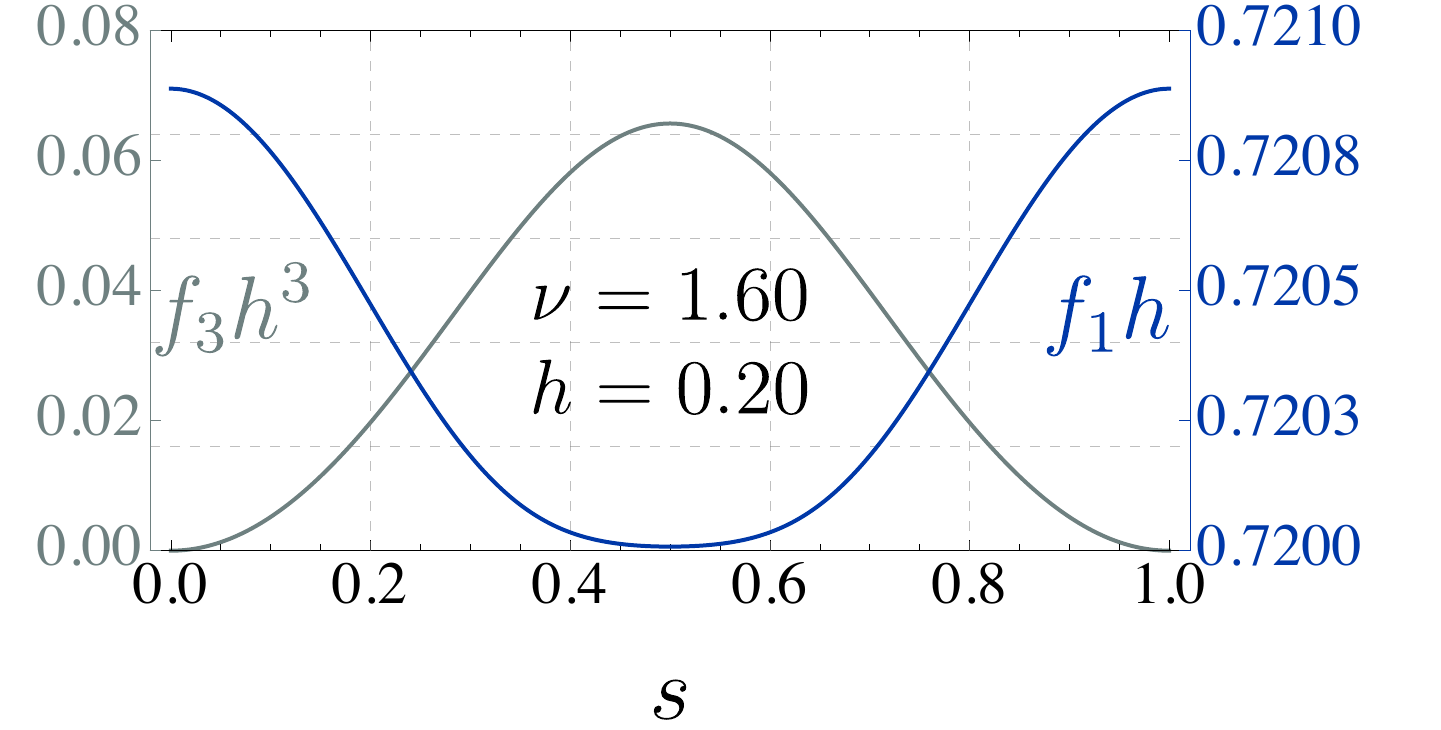}
		\caption{}\label{fig:energies_comparison_f}
	\end{subfigure}\hfill
	\begin{subfigure}[t]{0.30\linewidth}
		\centering
		\includegraphics[width=\linewidth]{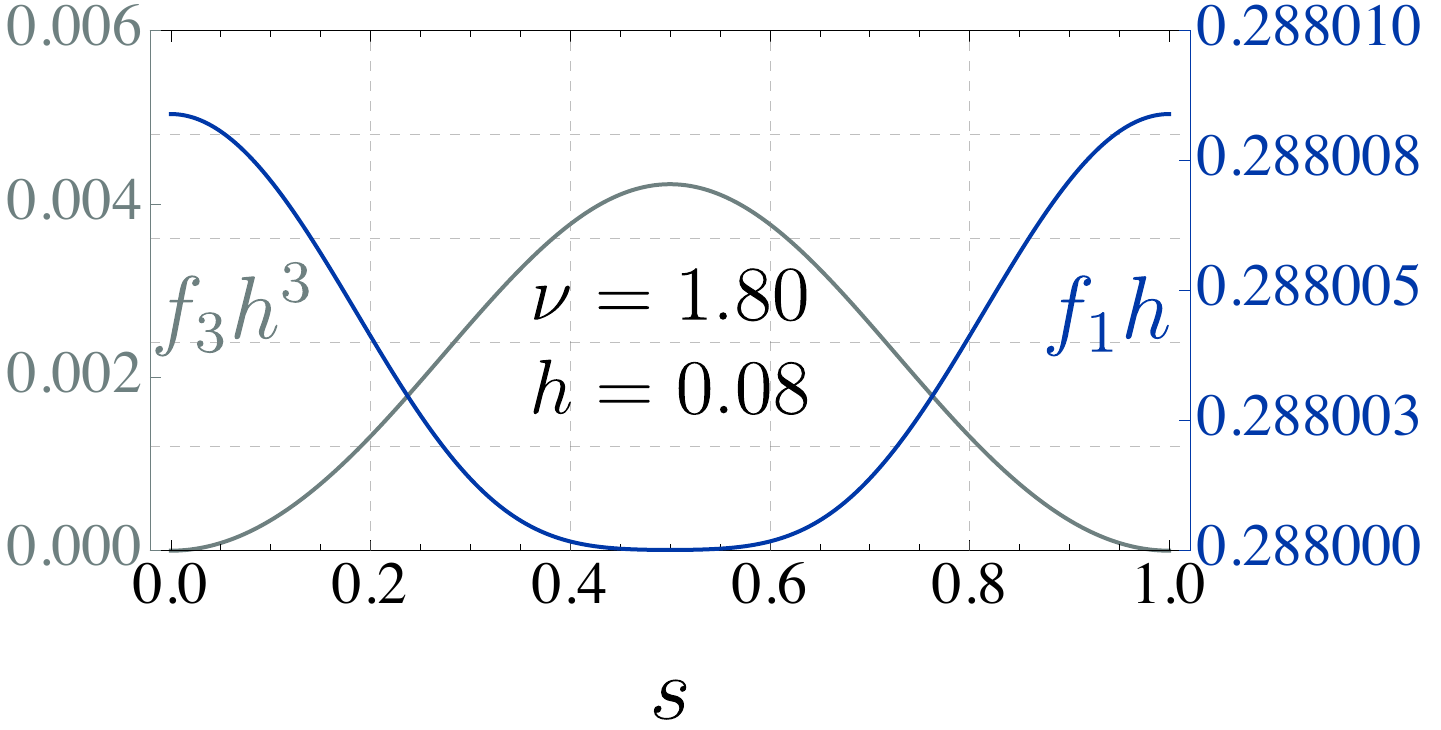}
		\caption{}\label{fig:energies_comparison_g}
	\end{subfigure}\quad
	\begin{subfigure}[t]{.30\linewidth}
		\centering
		\includegraphics[width=\textwidth]{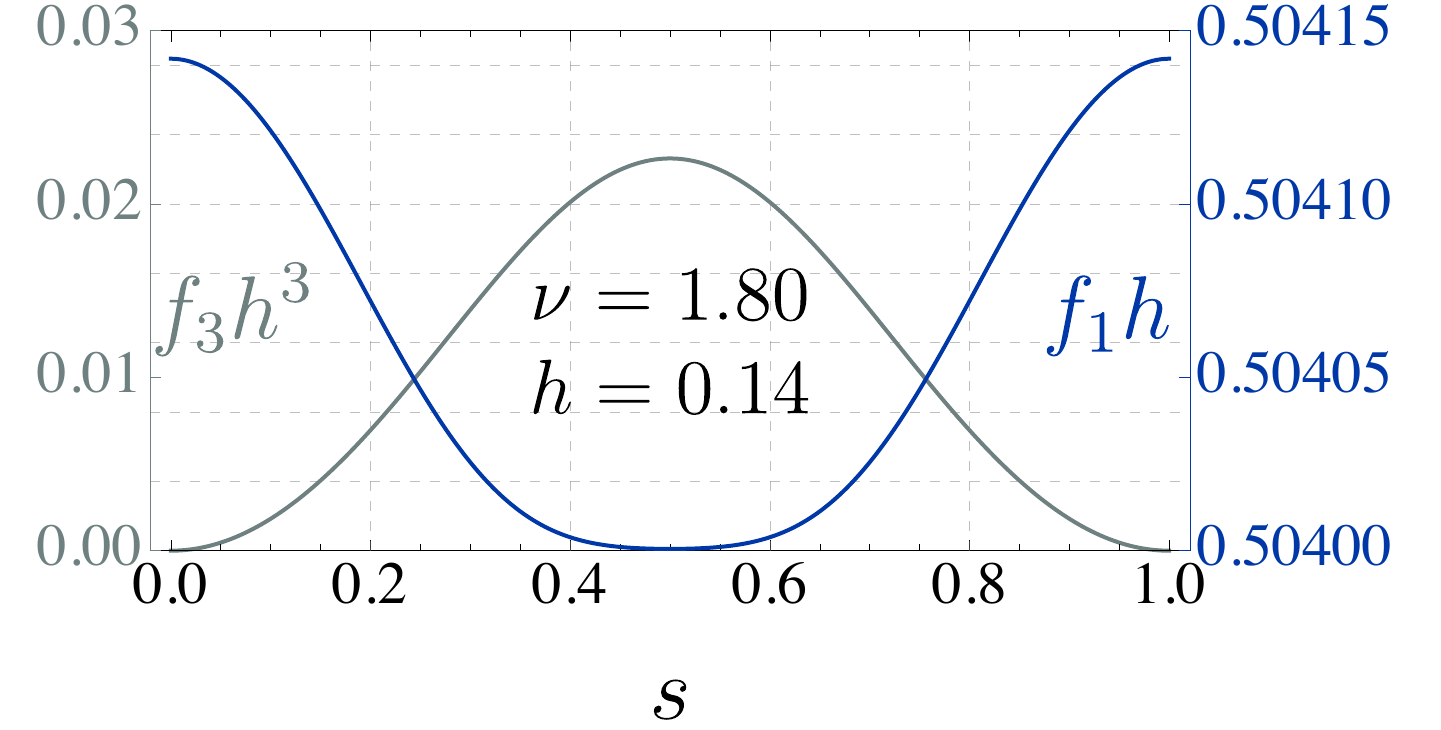}
		\caption{}\label{fig:energies_comparison_h}
	\end{subfigure}
	\begin{subfigure}[t]{.30\linewidth}
		\centering
		\includegraphics[width=\textwidth]{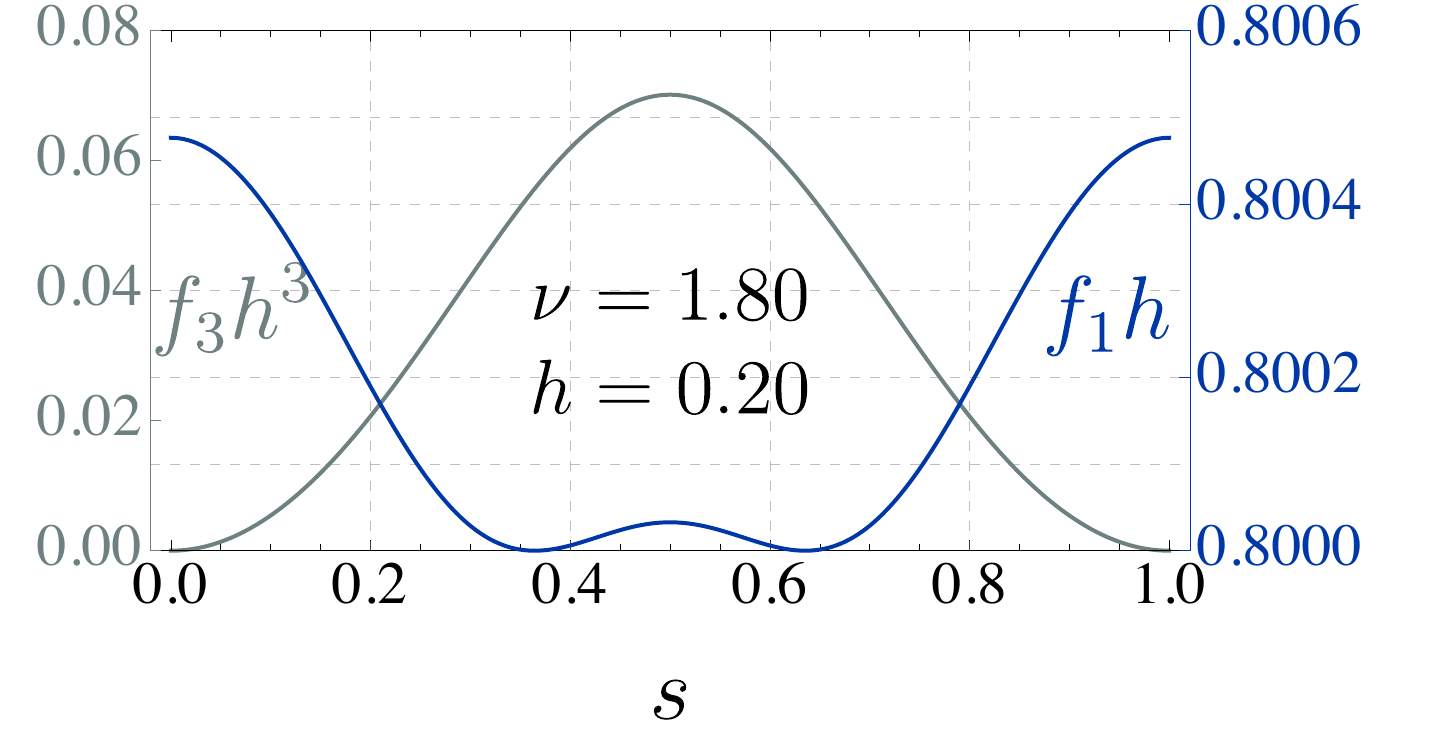}
		\caption{}\label{fig:energies_comparison_i}
	\end{subfigure}\hfill
	\begin{subfigure}[t]{0.30\linewidth}
		\centering
		\includegraphics[width=\linewidth]{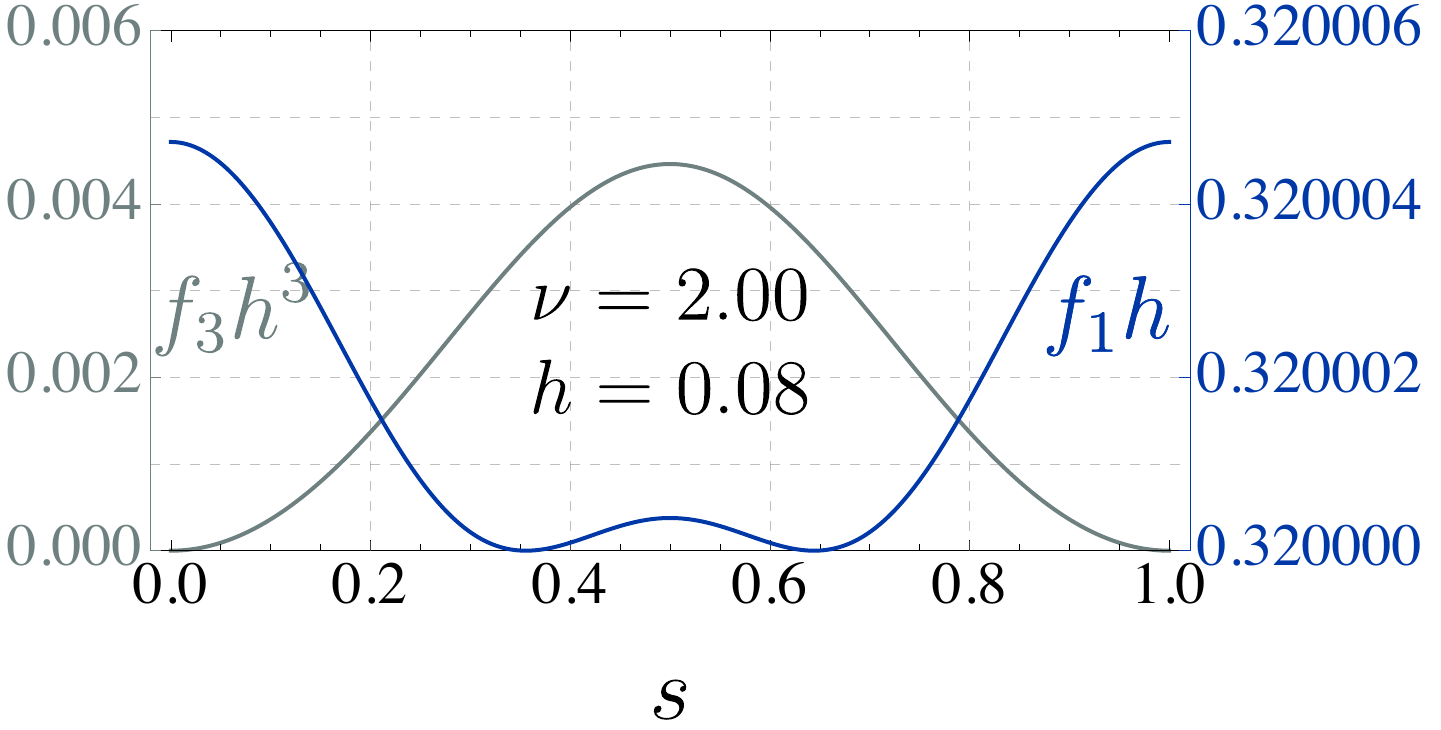}
		\caption{}\label{fig:energies_comparison_j}
	\end{subfigure}\quad
	\begin{subfigure}[t]{.30\linewidth}
		\centering
		\includegraphics[width=\textwidth]{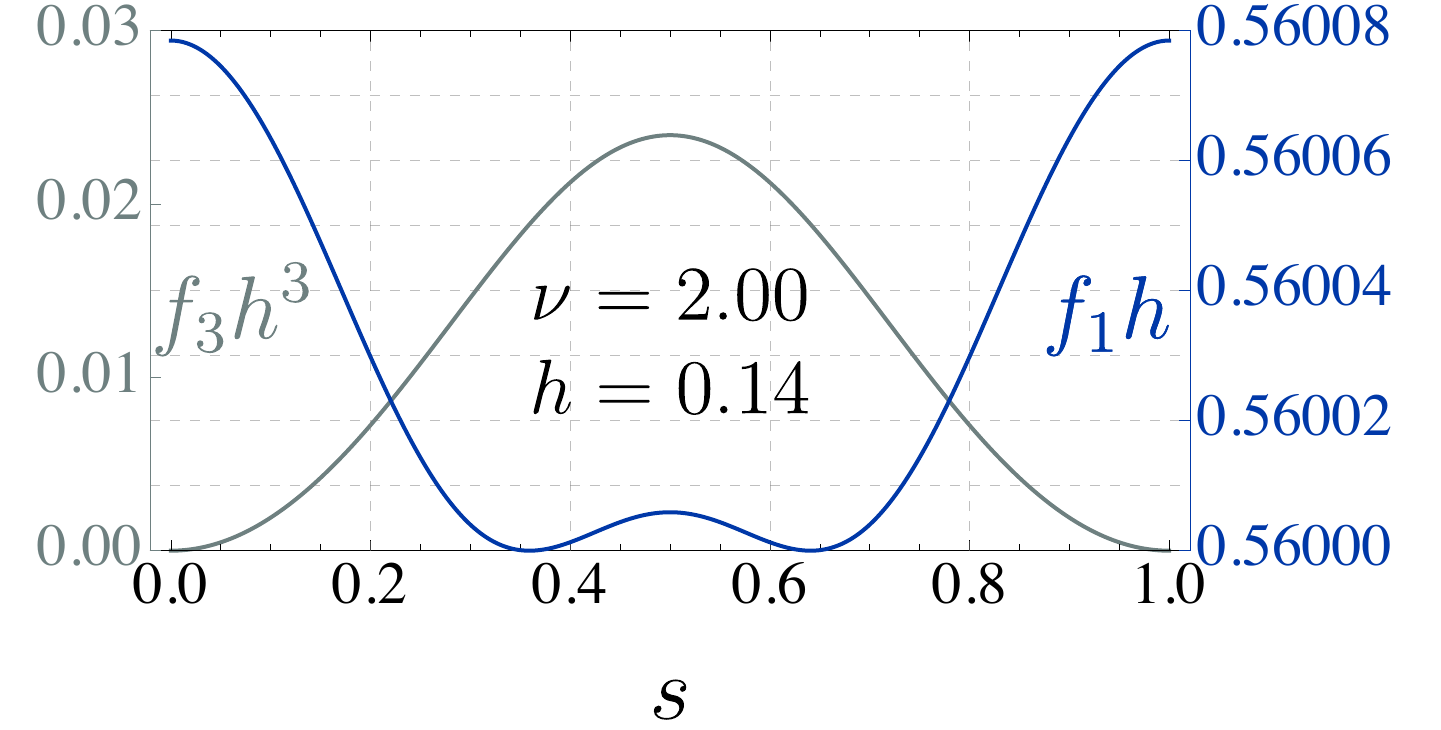}
		\caption{}\label{fig:energies_comparison_k}
	\end{subfigure}
	\begin{subfigure}[t]{.30\linewidth}
		\centering
		\includegraphics[width=\textwidth]{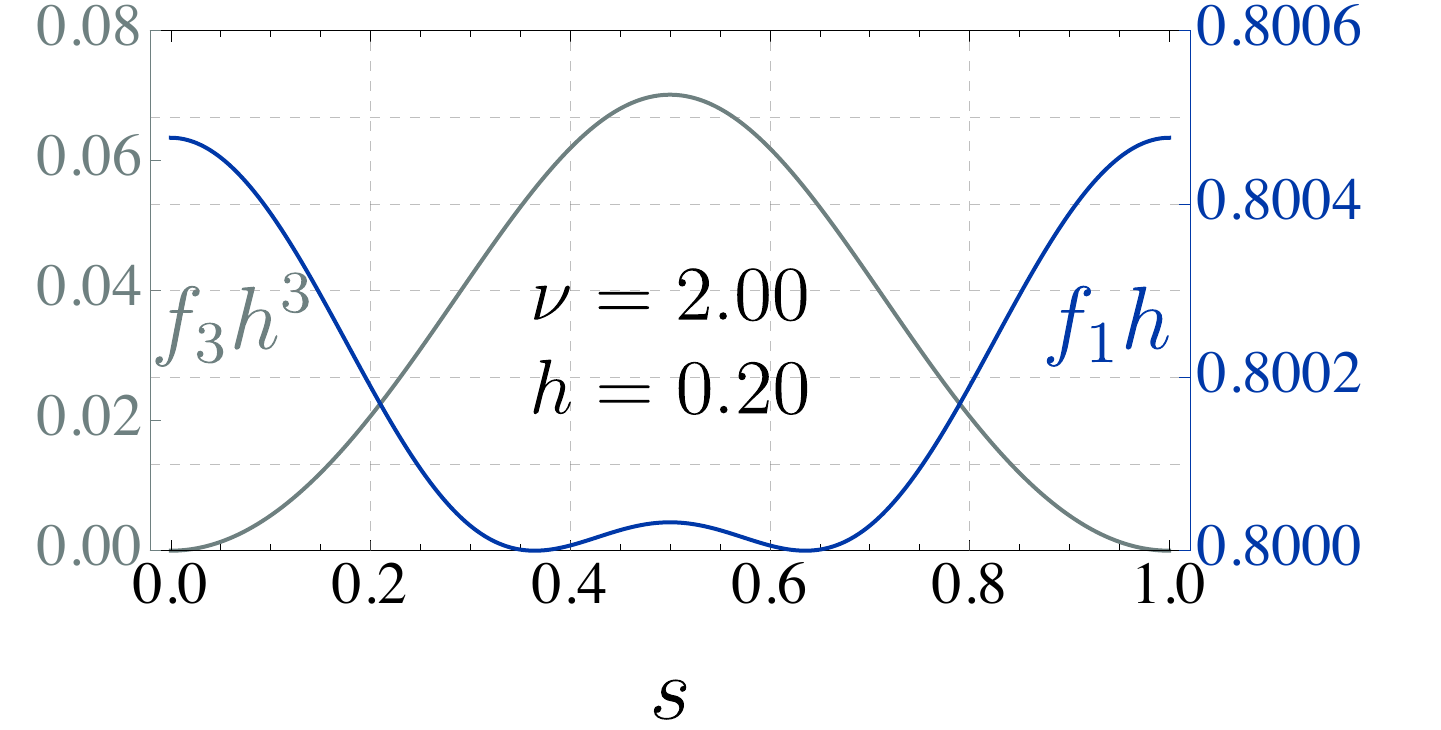}
		\caption{}\label{fig:energies_comparison_l}
	\end{subfigure}\hfill
	\caption{Several plots depicting the interplay of bending energy ($f_3h^3$) and stretching energy ($f_1h$) of a rectangular NPN ribbon for different values of $h$ and $\nu$ plotted against $s$. 
		In most panels, the two energies complement one another such that the local maximum of one corresponds to the local minimum of the other. This behavior slightly changes in panels (\ref{fig:energies_comparison_i}), (\ref{fig:energies_comparison_j}), (\ref{fig:energies_comparison_k}), and (\ref{fig:energies_comparison_l}), in that a new local maximum of the stretching energy forms in the center, where a local maximum of the bending energy lies.}
	\label{fig:energies_comparison}
\end{figure} 
Finally, we track the distribution of the bending energy vs the stretching energy along the arc-length coordinate in Fig.~\ref{fig:energies_comparison}.
The gray curve represents the bending energy, while the blue curve represents the stretching energy along the arc-length coordinate.
We observe that the stretching energy is always quantitatively dominant (by at least one order of magnitude) over the bending energy. However, there appears to be a sort of \emph{complementarity} relation between the two. Boundary conditions dictate that the bending energy vanishes at the ribbon's end-points, whereas the stretching energy can be freely distributed at equilibrium. Now, although dominant in magnitude, the latter chooses to be concentrated where the bending energy must be absent and to be minimum where the bending energy is maximum.
Thus, the stretching energy is primarily concentrated at the extreme ends of the ribbon, while the bending energy is concentrated in the middle. 
The stretching energy attains a local extremum at the center of the ribbon in all cases.
This extremum is the absolute minimum for most of the plots shown.
However, for plots in panels (\ref{fig:energies_comparison_i}), (\ref{fig:energies_comparison_j}), (\ref{fig:energies_comparison_k}), and (\ref{fig:energies_comparison_l}) this extremum morphs curiously into a local maximum.

\begin{figure}[h]
	\centering
	\begin{subfigure}[t]{0.30\linewidth}
		\centering
		\includegraphics[width=\linewidth]{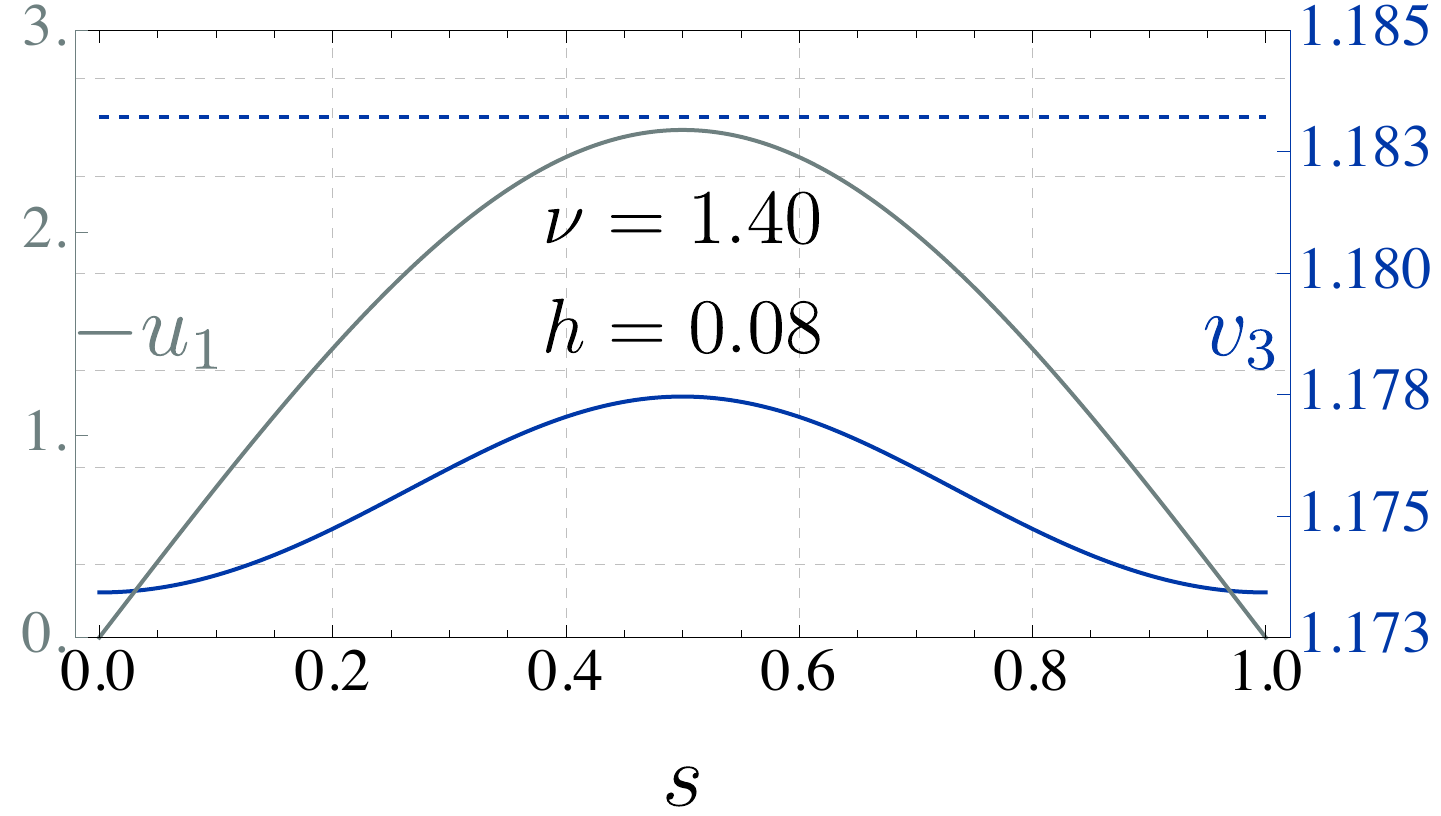}
		\caption{}\label{fig:sample_configuration_a}
	\end{subfigure}\quad
	\begin{subfigure}[t]{.30\linewidth}
		\centering
		\includegraphics[width=\textwidth]{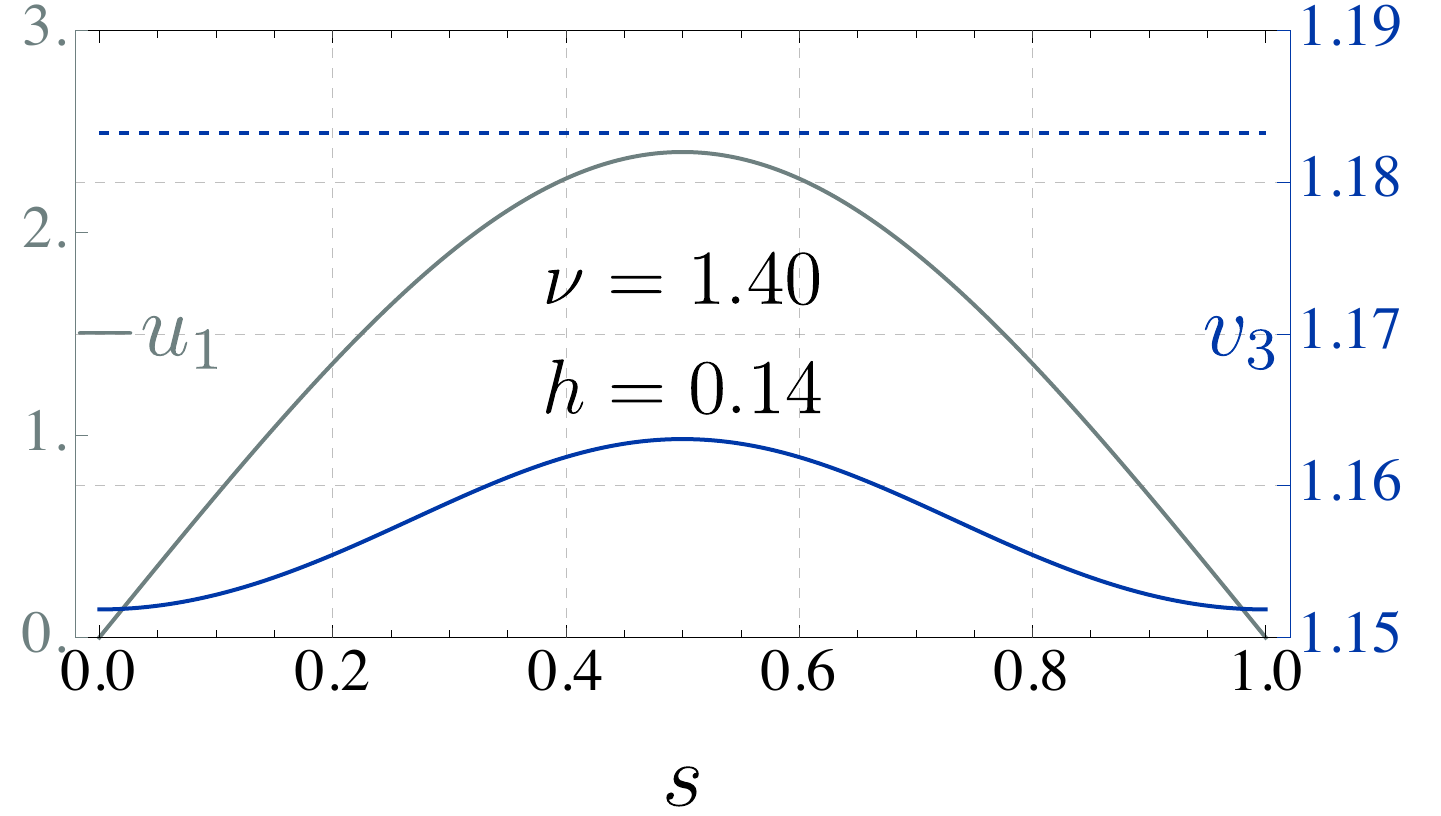}
		\caption{}\label{fig:sample_configuration_b}
	\end{subfigure}
	\begin{subfigure}[t]{.30\linewidth}
		\centering
		\includegraphics[width=\textwidth]{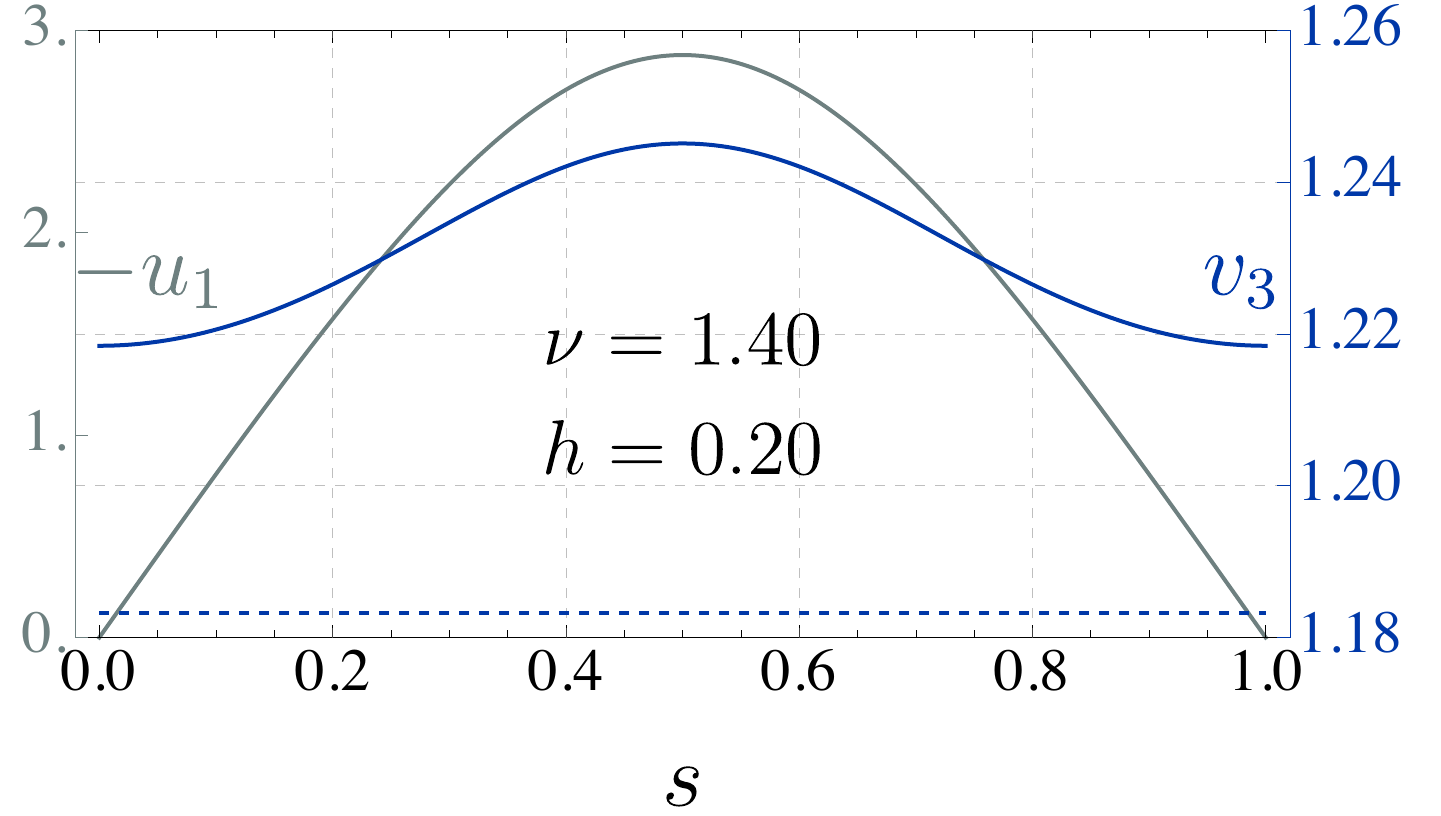}
		\caption{}\label{fig:sample_configuration_c}
	\end{subfigure}\hfill
	\begin{subfigure}[t]{0.30\linewidth}
		\centering
		\includegraphics[width=\linewidth]{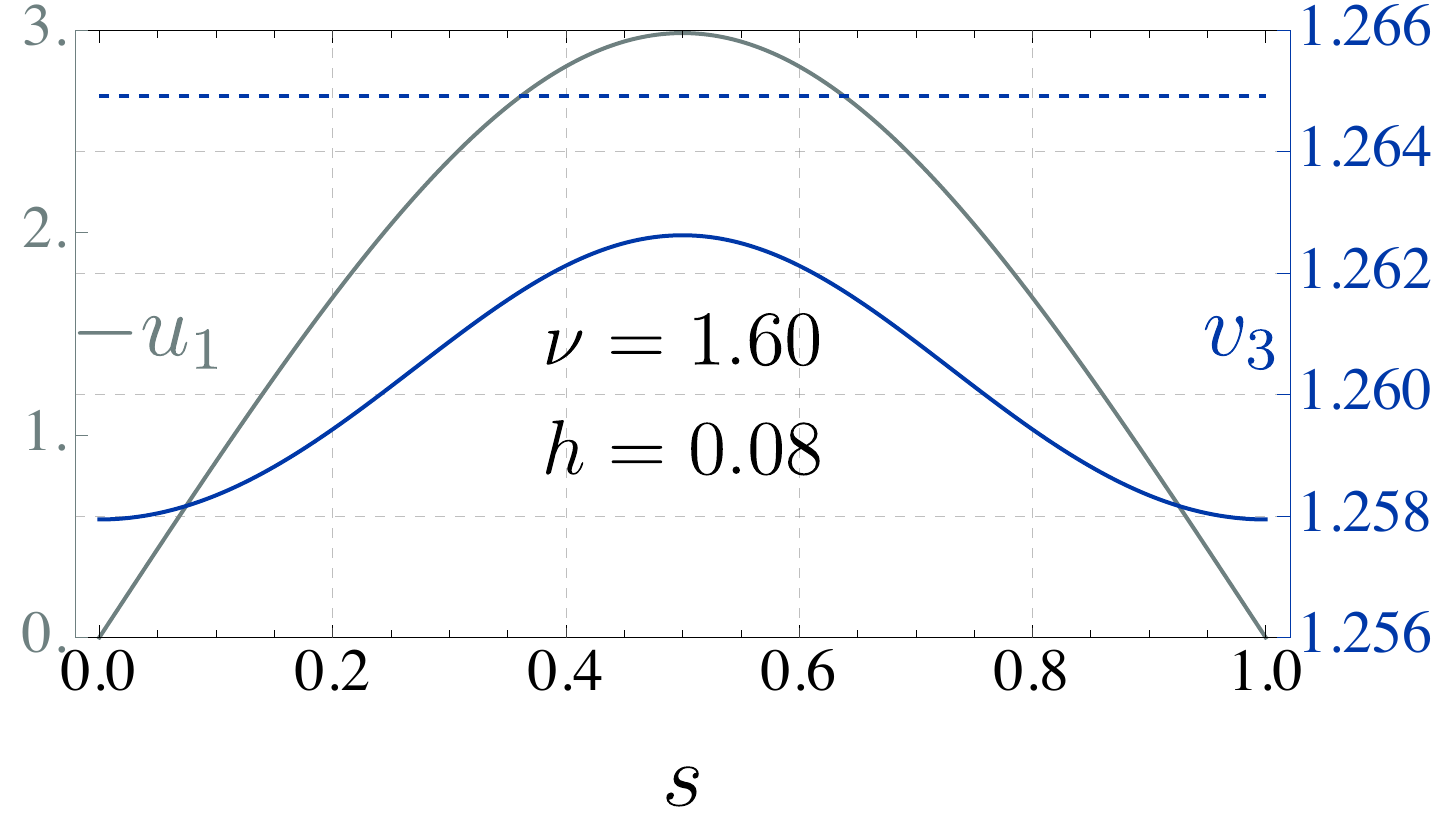}
		\caption{}\label{fig:sample_configuration_d}
	\end{subfigure}\quad
	\begin{subfigure}[t]{.30\linewidth}
		\centering
		\includegraphics[width=\textwidth]{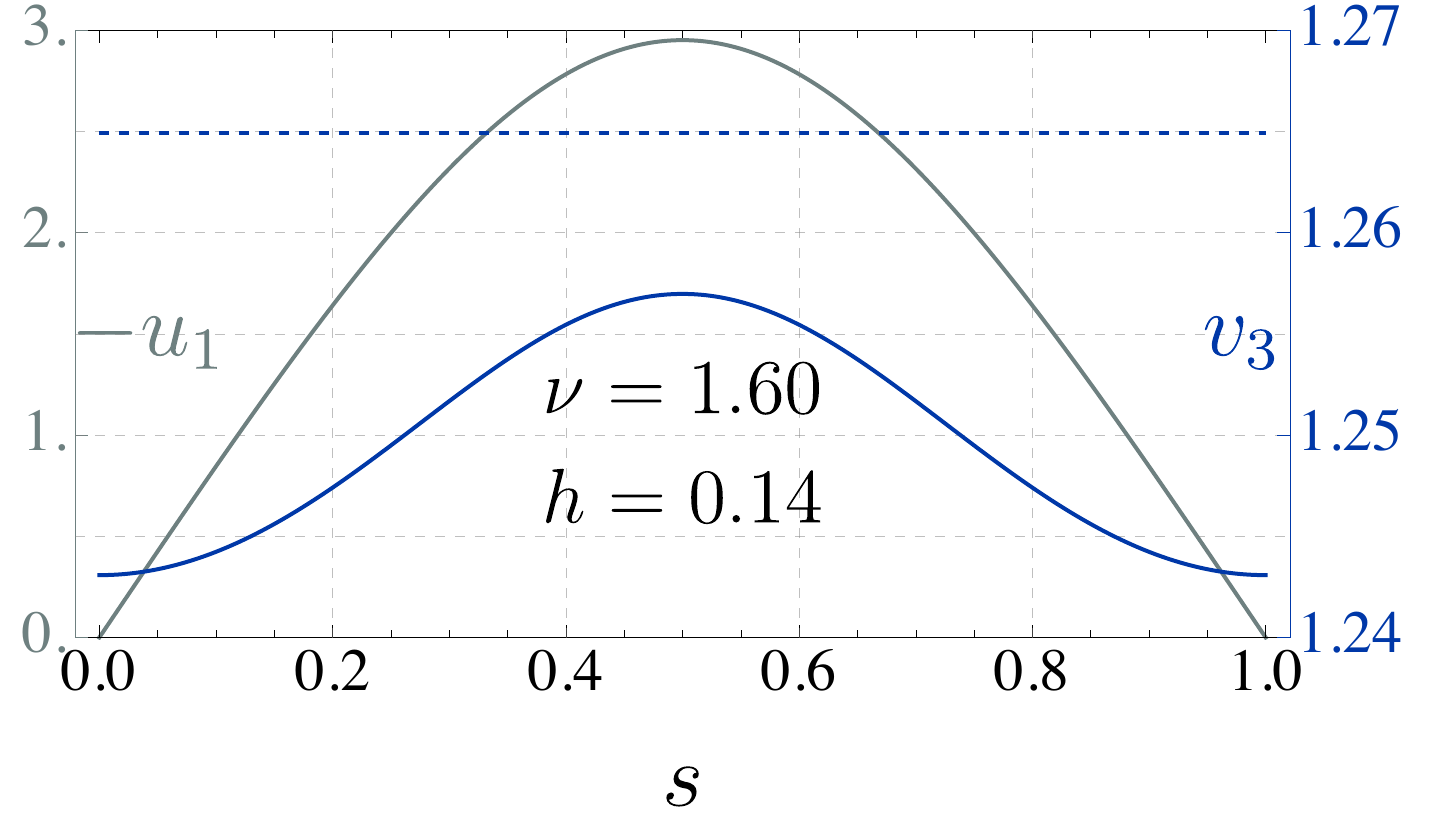}
		\caption{}\label{fig:sample_configuration_e}
	\end{subfigure}
	\begin{subfigure}[t]{.30\linewidth}
		\centering
		\includegraphics[width=\textwidth]{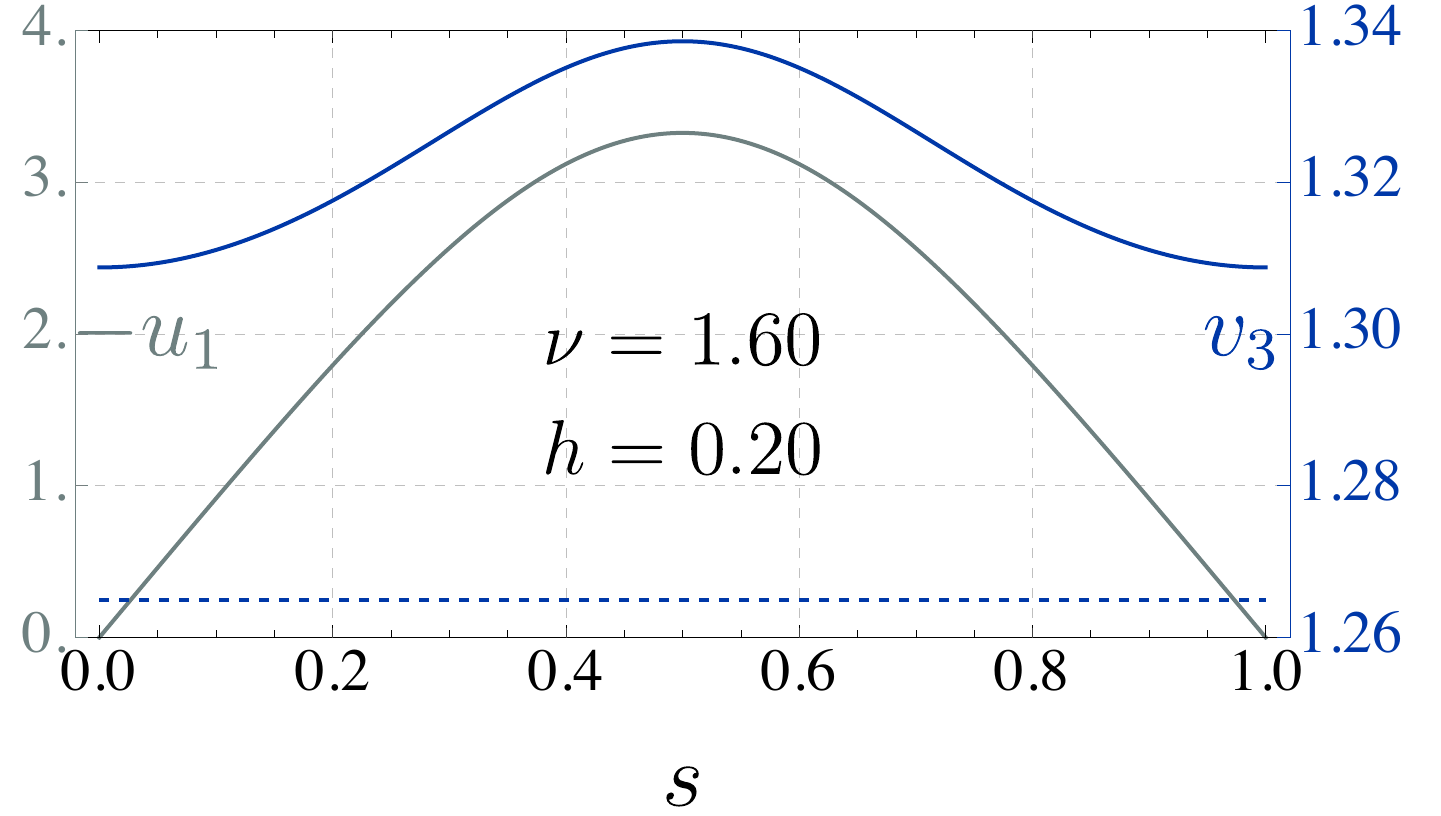}
		\caption{}\label{fig:sample_configuration_f}
	\end{subfigure}\hfill
	\begin{subfigure}[t]{0.30\linewidth}
		\centering
		\includegraphics[width=\linewidth]{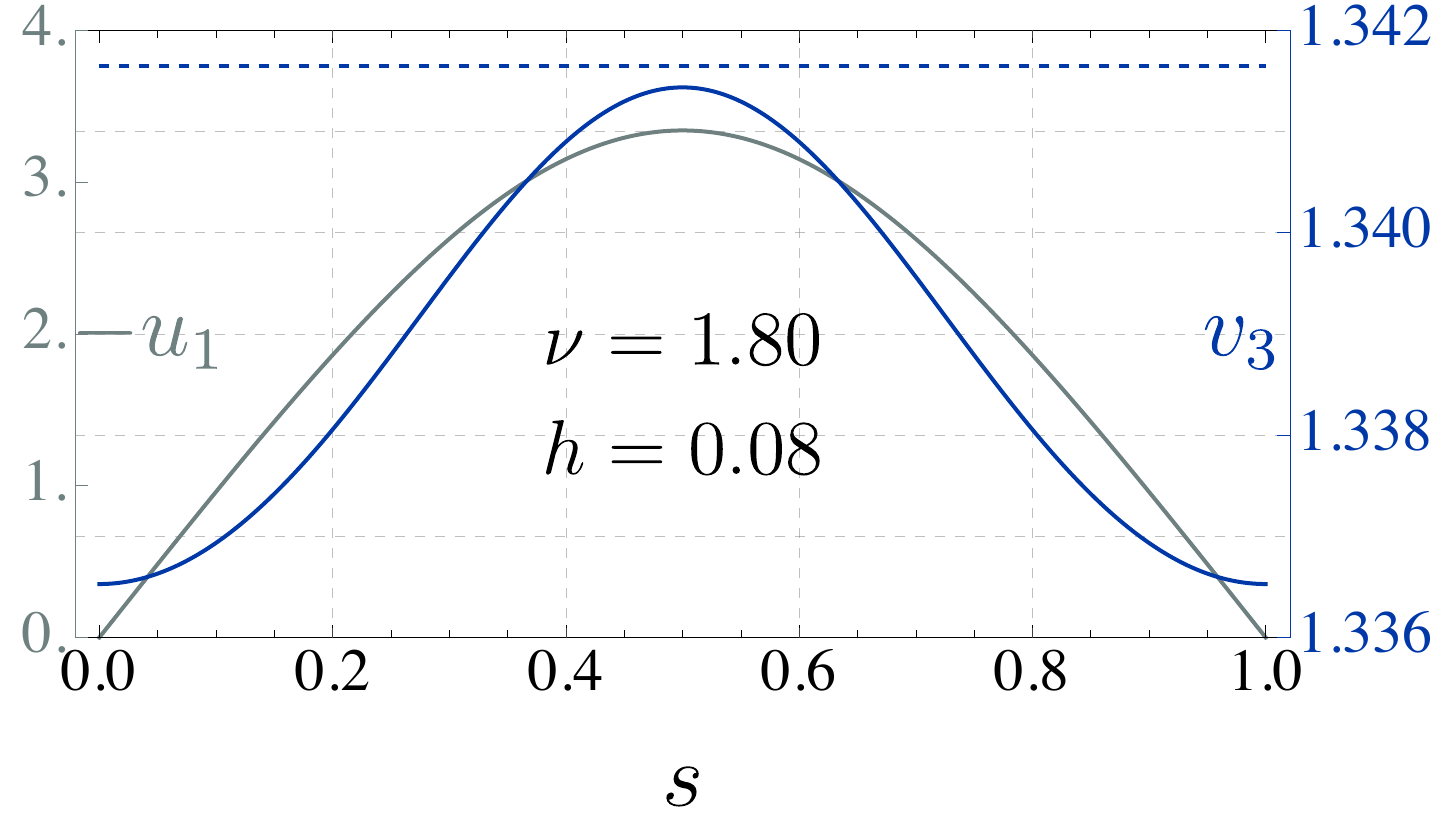}
		\caption{}\label{fig:sample_configuration_g}
	\end{subfigure}\quad
	\begin{subfigure}[t]{.30\linewidth}
		\centering
		\includegraphics[width=\textwidth]{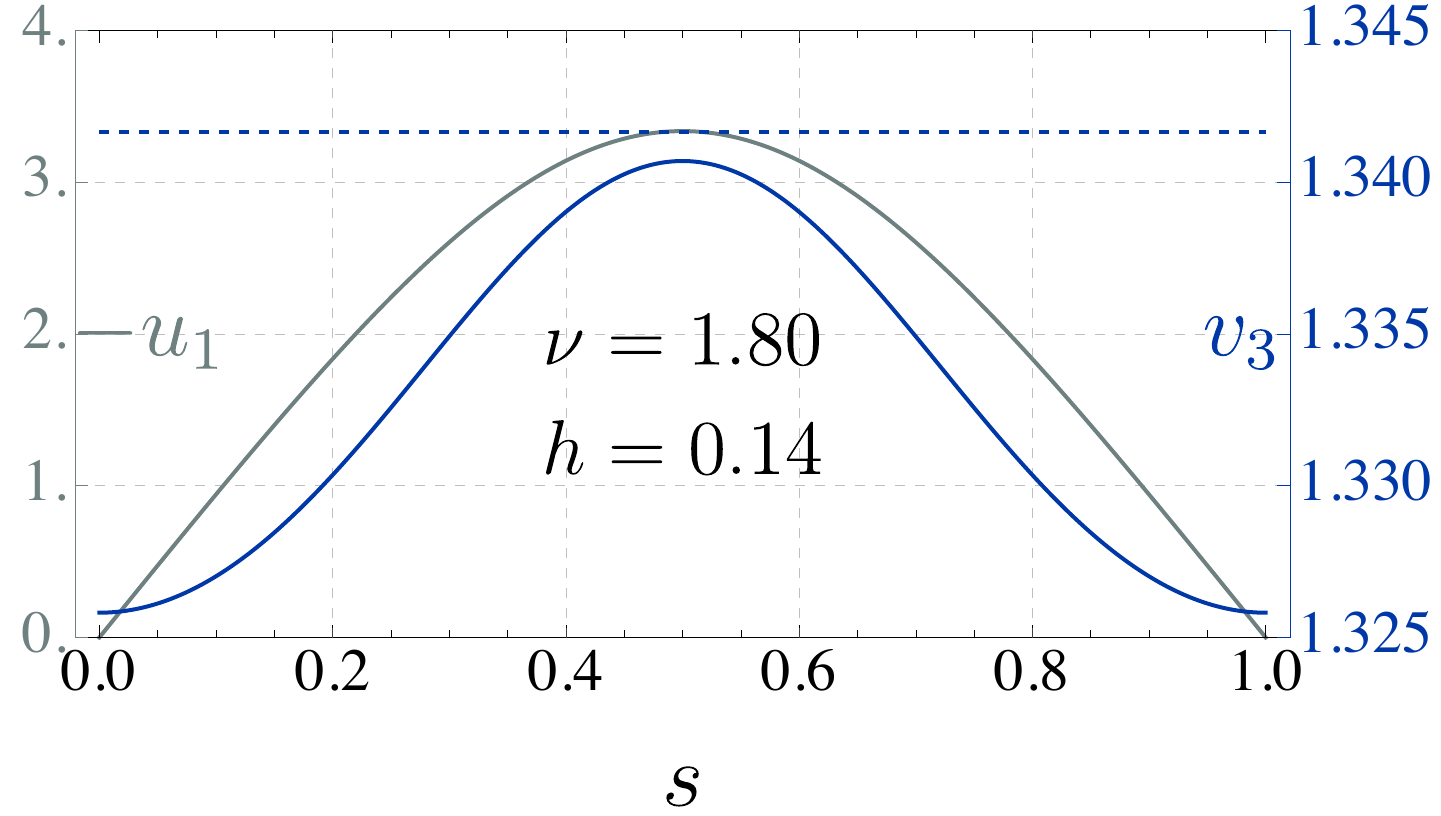}
		\caption{}\label{fig:sample_configuration_h}
	\end{subfigure}
	\begin{subfigure}[t]{.30\linewidth}
		\centering
		\includegraphics[width=\textwidth]{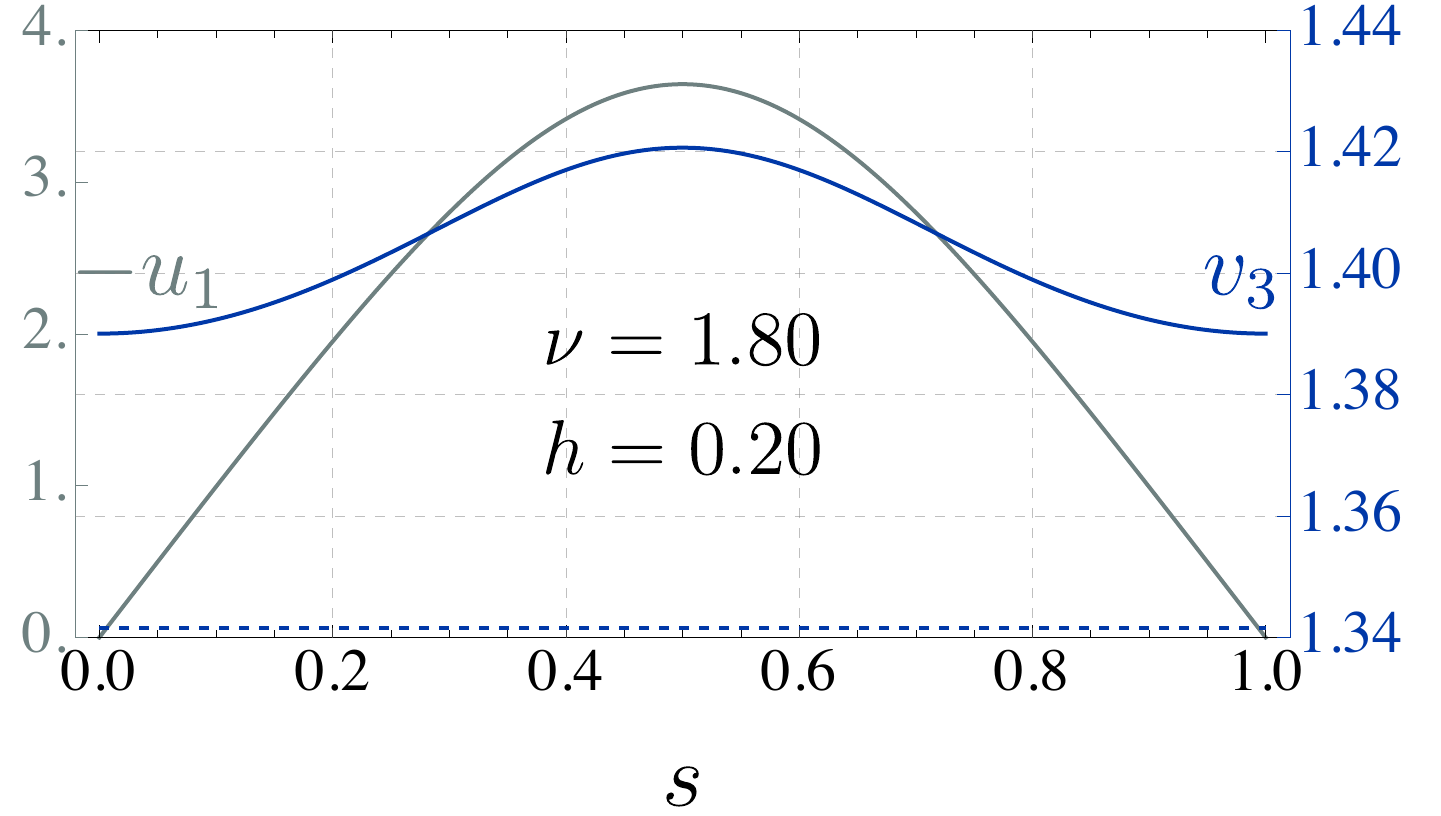}
		\caption{}\label{fig:sample_configuration_i}
	\end{subfigure}\hfill
	\begin{subfigure}[t]{0.30\linewidth}
		\centering
		\includegraphics[width=\linewidth]{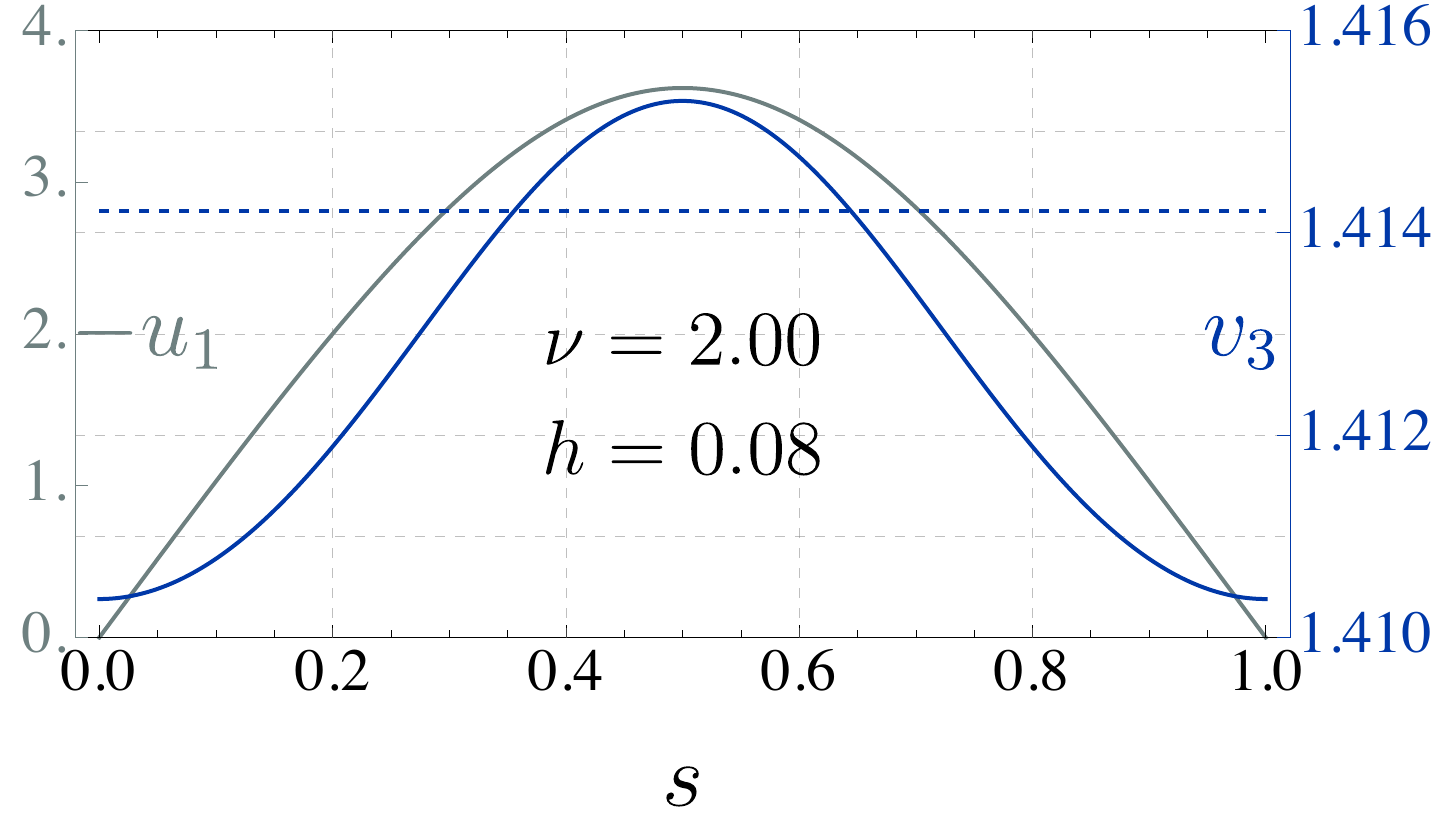}
		\caption{}\label{fig:sample_configuration_j}
	\end{subfigure}\quad
	\begin{subfigure}[t]{.30\linewidth}
		\centering
		\includegraphics[width=\textwidth]{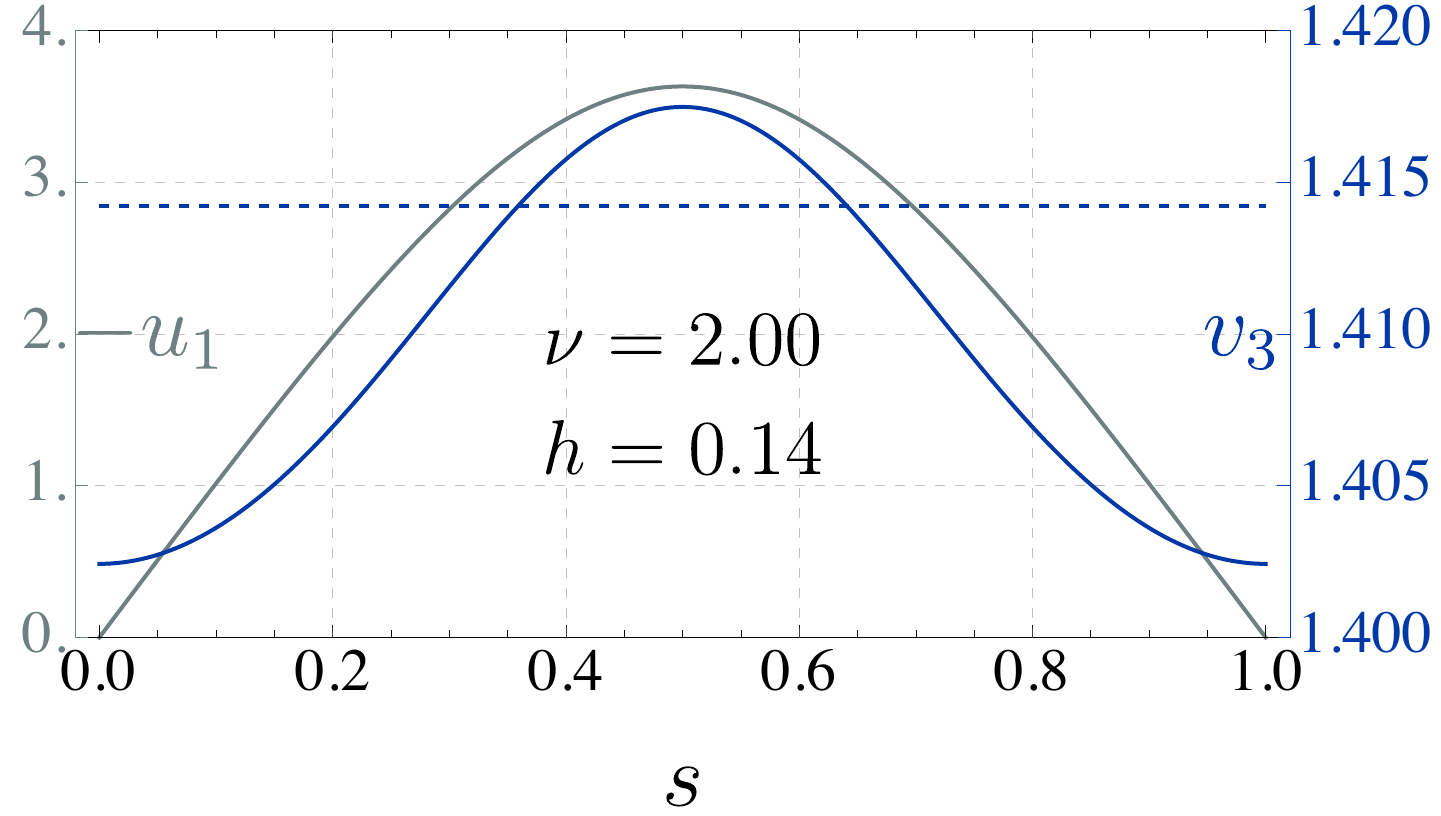}
		\caption{}\label{fig:sample_configuration_k}
	\end{subfigure}
	\begin{subfigure}[t]{.30\linewidth}
		\centering
		\includegraphics[width=\textwidth]{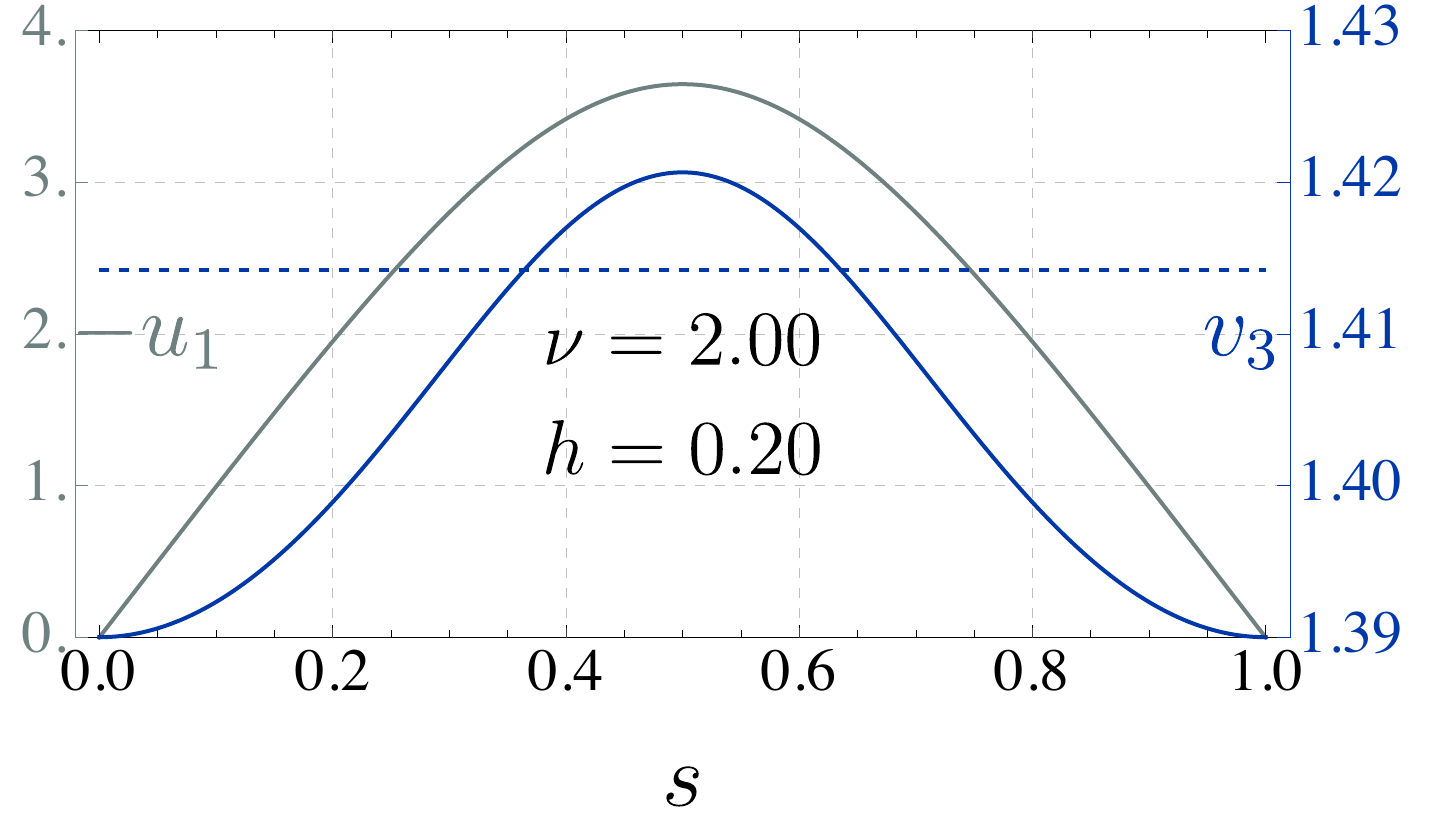}
		\caption{}\label{fig:sample_configuration_l}
	\end{subfigure}\hfill
	\caption{Several plots showing the distribution of the curvature $-u_1>0$ and the stretch $v_3$ of a rectangular NPN ribbon for different values of $h$ and $\nu$ against $s$. The location of the maxima and minima of the curvature and the stretch on the centerline coincides for all cases. The values of $v_3$ hover close to the local minimum at $v_3^*=\sqrt{\nu}$ of the stretching energy function $f_1$ in \eqref{eq:energy_densities_rectangular}.}
	\label{fig:sample_configuration}
\end{figure}
On similar lines, we plot in Fig.~\ref{fig:sample_configuration} the stretch $v_3$ and the curvature $-u_1$ for various values of $\nu$ and $h$.
We observe that the curvature and the stretch both attain their respective maxima at the center of the ribbon, with their minima located at the two ends. The coincidence of maximum curvature and maximum stretch was perhaps to be expected, as the bending rigidity in \eqref{eq:energy_densities_rectangular} decreases with increasing $v_3$.
We also observe that the variation in the stretch along the length is relatively much smaller, and $v_3$ does not differ much from  the value $v_3^*=\sqrt{\nu}$, shown by the dashed line on each plot, at which the function $f_1$ in \eqref{eq:energy_densities_rectangular} attains its minimum.
The stretching measure $v_3$ is moderate (corresponding to a length increment of nearly $20\%$) for small $\nu$ and significant (corresponding to a length increment of nearly $40\%$) for large $\nu$.

\begin{figure}[h!]
	\centering
	\includegraphics[width=.75\textwidth]{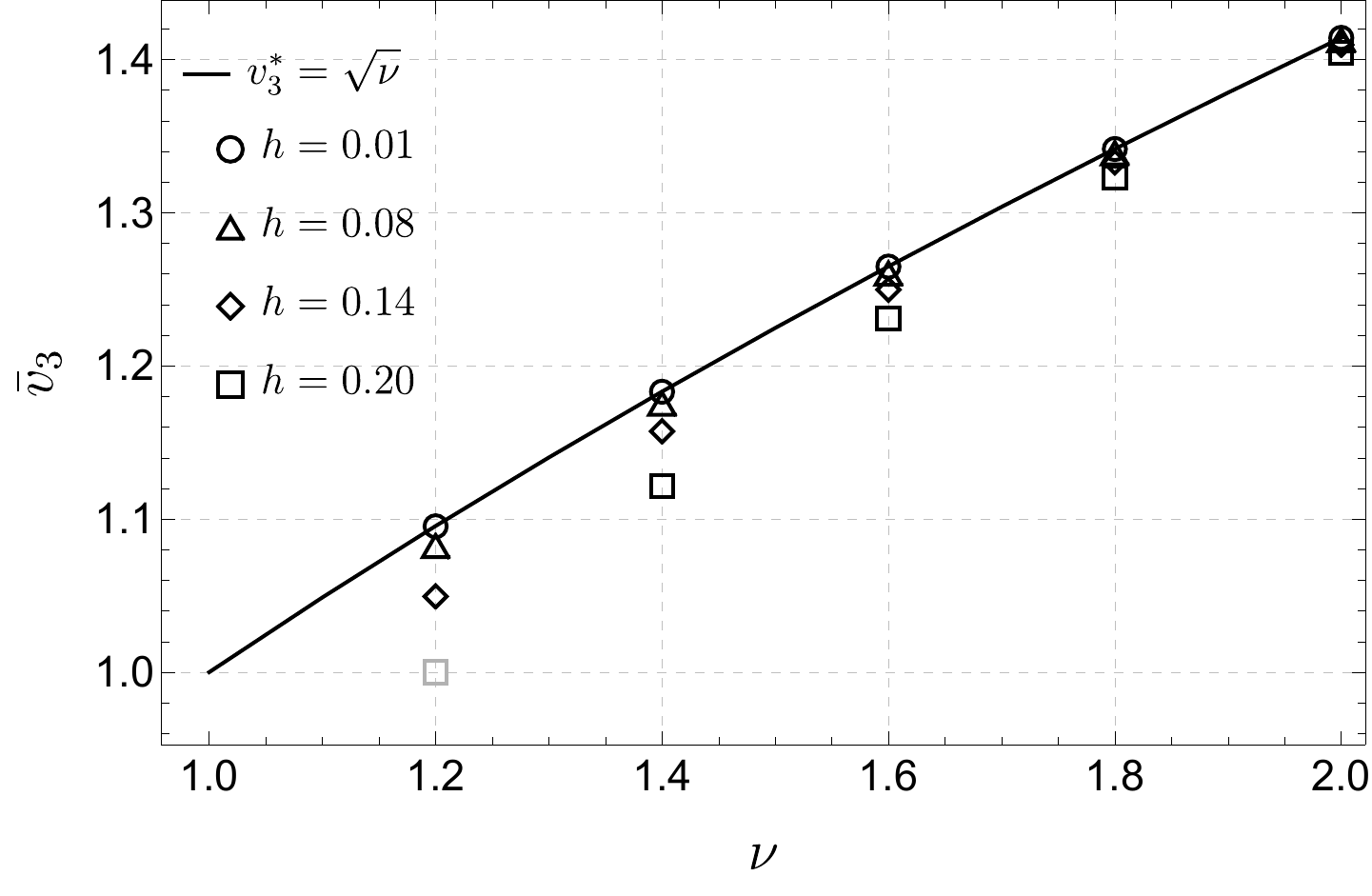}
	\caption{The values of the average stretch $\bar{v}_3$, as defined by \eqref{eq:average_v_3}, are compared with the minimum $v_3^\ast$ of the stretching energy for the different values of $h$ and $\nu$ considered in our analysis. The gray datapoint ``$\color{gray}{\square}$''  has been added for $\nu=1.2$ at $\bar{v}_3=1$, corresponding to the trivial solution in \eqref{eq:trivial_solution}, which, as shown in Fig.~\ref{fig:bifurcation_diagrams_d}, is the only energy minimizer for that value of $\nu$.}
	\label{fig:average_stretch}	
\end{figure}
We see from Fig.~\ref{fig:average_stretch} that, for given $\nu$, an increase in $h$  results into a decrease in the average stretch defined by 
\begin{equation}
	\label{eq:average_v_3}
	\bar{v}_3:=\int_0^1v_3\,ds.
\end{equation}
Comparing this latter with $v^\ast_3$, we see that $\bar{v}_3\le v^\ast_3$ for all values of $h$ and $\nu$ that  we explored.  Moreover, for given $h$, $\bar{v}_3$ and $v^\ast_3$ get closer as $\nu$ increases and, for given $\nu$, they become farther apart as $h$ increases (see Fig.~\ref{fig:average_stretch}). Another telling feature is revealed by Fig.~\ref{fig:average_stretch}: for the smallest value of $h$, $\bar{v}_3$ falls nicely over $v_3^\ast$, marking the expected prevalence of a membrane-like behavior, which is soon lost as $h$ is increased.

Both Fig.~\ref{fig:average_stretch} and the plots in Figs.~\ref{fig:sample_configuration_j}, \ref{fig:sample_configuration_k}, and \ref{fig:sample_configuration_l} show clearly that, for large values of $\nu$, $h$ has little effect on the deformation of the ribbon, in complete accordance with Fig.~\ref{fig:varying_thickness}. We shall designate this as the \emph{bleaching} regime. Our analysis  justifies approximating $v_3$ with $v^\ast_3$ in the bleaching regime: there the membrane-like behavior prevails \emph{irrespective} of the value of $h$ and the bending energy acts as a \emph{selection criterion} that singles out a deformation out of myriads with the same minimum stretching energy. This, however,  is no longer the case  for smaller values of the activation parameter (say, near $\nu=1$), where the independent measures of deformation $(v_3,u_1)$ are closely interlaced and the mechanical behavior of the ribbon cannot be classified neither as ``pure-membrane'' nor as ``pure-plate''.   

\section{Conclusions}\label{sec:conclusions}
In this paper, which is the ideal companion of \cite{singh:model}, we studied the out-of-plane deformations of a narrow ribbon consisting of a nematic polymer network, a nematic elastomer where the degree of cross-linking is sufficiently high to justify the enslavement of the nematic director to the polymer matrix deformation.

The material is  activated by a change in temperature: this produces a mismatch, measured by a control parameter $\nu>1$, between the degrees of order in the polymer chains organization in the reference and current configurations. This drives a spontaneous deformation of the ribbon. In our theory, both stretching and bending energies are contemplated, scaling with different powers of the ribbon's thickness $h$. 

We deliberately kept simple the deformation pattern, choosing boundary conditions compatible with the absence of twist, as our primary aim was exploring the interplay between stretching and bending components of the elastic energy while both $\nu$ and $h$ were independently varied.

Two major conclusions were reached here. First, stretching and bending energies appear to obey a \emph{complementarity} relation: one is maximum where the other is minimum. Second, our findings point to the existence of a \emph{bleaching} regime: for $\nu$ sufficiently large, the deformation of the ribbon is essentially independent of $h$. This latter feature may be relevant to applications where displacements induced by activation are desired to be large in not too thin ribbons.

Clearly, envisioning twist in the admissible class of deformations  might have a bearing on our present conclusions, but not---we believe---to the point of obliterating them. It would be desirable to address this issue in the future.

\backmatter

\bmhead{Acknowledgments}
H.S. is grateful to S.P.C. Dhanakoti and T. Yu for helpful discussions regarding \texttt{AUTO-07P}.
He also acknowledges partial support by Swiss National Science Foundation Grant $200020\_182184$ to J.H. Maddocks.

\section*{Declarations}
\subsection*{Competing interests}
The authors have no conflicts of interest to declare that are relevant to the content of this article.

\begin{appendices}

\section{Minimizing stretch}\label{sec:stretching}
Here we show that the stretching content $f_\mathrm{s}(\C)$ in \eqref{eq:f_s} has precisely one stationary point $\C_0$ subject to the inextensibility constraint 
\begin{equation}
	\label{eq:det_C=1}
	\det\C=1
\end{equation}
generated by \eqref{eq:det_nabla_y}.

Since 
\begin{equation}
	\label{eq:determinant_derivative}
	\frac{\partial}{\partial\C}\det\C=(\det\C)(\C\trans)^{-1},
\end{equation}
see for example \cite[p.\,24]{gurtin:mechanics}, by \eqref{eq:det_C=1} and the symmetry of $\C$, it follows from \eqref{eq:f_s} that the stationary condition 
\begin{equation}
	\label{eq:stationary_condition}
	\frac{\partial f_\mathrm{s}}{\partial\C}=\lambda\frac{\partial}{\partial\C}\det\C
\end{equation}
is equivalently written as 
\begin{equation}
	\label{eq:stationary_condition_equivalent}
	\mathbf{I}_2+\left(S_0-\frac{S}{(\n_0\cdot\C\n_0)^2}\right)\n_0\otimes\n_0=\lambda\C^{-1},
\end{equation}
where $\mathbf{I}_2$ is the two-dimensional identity and $\lambda$ is the Lagrange multiplier associated with \eqref{eq:det_C=1}.

Acting with $\C$ on both sides of \eqref{eq:stationary_condition_equivalent}, we arrive at 
\begin{equation}
	\label{eq:stationarity_condition_final}
	\C+\left(S_0-\frac{S}{(\n_0\cdot\C\n_0)^2}\right)\C\n_0\otimes\n_0=\lambda\mathbf{I}_2,
\end{equation}
which easily implies that a unit vector $\n_0^\perp$ perpendicular to $\n_0$ in the $\{\e_3,\e_1\}$ plan is an eigenvector of $\C$ with eigenvalue $\lambda$, while $\n_0$ is an eigenvector with eigenvalue $1/\lambda$, by \eqref{eq:det_C=1}. Taking the inner product with  $\n_0\otimes\n_0$ of both sides of \eqref{eq:stationarity_condition_final}, we readily arrive at 
\begin{equation}
	\label{eq:lambda_formula}
	\lambda=\nu,
\end{equation}
where $\nu$ is as in \eqref{eq:nu_definition}, and we write the minimizer $\C_0$ of $f_\mathrm{s}$ as\footnote{That the \emph{only} stationary point $\C_0$ actually minimizes $f_\mathrm{s}$ follows easily from being $f_\mathrm{s}$ bounded below (but not above).}
\begin{equation}
	\label{eq:C_0}
	\C_0=\frac{1}{\nu}\n_0\otimes\n_0+\nu\n_0^\perp\otimes\n_0^\perp.
\end{equation}
Clearly, $\C_0$ reduces to $\mathbf{I}_2$ for $\nu=1$ (that is, for $S=S_0$).




\end{appendices}





\end{document}